\documentclass[aps,showpacs,onecolumn,
superscriptaddress]{revtex4}
\usepackage{graphicx}
\usepackage{dcolumn}
\usepackage{bm}
\usepackage{color}
\usepackage[normalem]{ulem}
\usepackage[dvipsnames]{xcolor}
\usepackage{hyperref}
\hypersetup{colorlinks=true, linkcolor=blue, citecolor=cyan}
\usepackage{mathrsfs}
\usepackage[utf8]{inputenc}
\usepackage{mathtools}
\usepackage{doi}
\usepackage{amsmath}
\usepackage{amssymb}
\usepackage{multirow}

\providecommand{\aap}{A\&A}

\providecommand{\apj}{ApJ}
\providecommand{\apjl}{ApJL}
\providecommand{\araa}{ARA\&A}
\providecommand{\jcap}{JCAP}
\providecommand{\mnras}{MNRAS}
\providecommand{\nat}{Nature}
\providecommand{\pasj}{PASJ}
\providecommand{\pasp}{PASP}
\providecommand{\physrep}{Phys.~Rep.}
\providecommand{\prd}{Phys.~Rev.~D}
\providecommand{\prl}{Phys.~Rev.~Lett.}
\usepackage{scalerel}
\usepackage{tikz}
\usetikzlibrary{svg.path}
\definecolor{orcidlogocol}{HTML}{A6CE39}
\tikzset{
  orcidlogo/.pic={
    \fill[orcidlogocol] svg{M256,128c0,70.7-57.3,128-128,128C57.3,256,0,198.7,0,128C0,57.3,57.3,0,128,0C198.7,0,256,57.3,256,128z};
    \fill[white] svg{M86.3,186.2H70.9V79.1h15.4v48.4V186.2z}
                 svg{M108.9,79.1h41.6c39.6,0,57,28.3,57,53.6c0,27.5-21.5,53.6-56.8,53.6h-41.8V79.1z M124.3,172.4h24.5c34.9,0,42.9-26.5,42.9-39.7c0-21.5-13.7-39.7-43.7-39.7h-23.7V172.4z}
                 svg{M88.7,56.8c0,5.5-4.5,10.1-10.1,10.1c-5.6,0-10.1-4.6-10.1-10.1c0-5.6,4.5-10.1,10.1-10.1C84.2,46.7,88.7,51.3,88.7,56.8z};
  }
}

\newcommand\orcidicon[1]{\href{https://orcid.org/#1}{\mbox{\scalerel*{
\begin{tikzpicture}[yscale=-1,transform shape]
\pic{orcidlogo};
\end{tikzpicture}
}{|}}}}

\begin{document}
\title{Probing black holes surrounded by Dust field in Quantum Fluctuation Modified Gravity with twin-peak QPOs}

\author{Kamola~Shermatova~\orcidicon{0009-0002-1586-164X}}
\email{sh.kamola95@gmail.com}
\affiliation{National University of Uzbekistan, University Str. 4, Tashkent 100174, Uzbekistan}

\author{Qodir~Badalov\orcidicon{0009-0000-2758-3777}}
\email{kr.badalov@samdpi.uz}
\affiliation{Samarqand State Pedagogical Institute, Spitamen Shokh Street 166, Samarkand 140100, Uzbekistan}

\author{Abubakir~Shermatov\orcidicon{0009-0009-4044-4507}} 
\email{shermatov.abubakir98@gmail.com}
\affiliation{Institute of Fundamental and Applied Research, National Research University TIIAME, Kori Niyoziy 39, Tashkent 100000, Uzbekistan}
\affiliation{Tashkent State Technical University, Tashkent 100095, Uzbekistan}

\author{Javlon~Rayimbaev\orcidicon{0000-0001-9293-1838}}
\email{javlon@astrin.uz}
\affiliation{School of Physics, Harbin Institute of Technology, Harbin 150001, People’s Republic of China}
\affiliation{University of Tashkent for Applied Sciences, Str. Gavhar 1, Tashkent 100149, Uzbekistan}
\affiliation{Institute of Theoretical Physics, National University of Uzbekistan, University Str. 4, Tashkent 100174, Uzbekistan}

\author{Bekzod~Rahmatov\orcidicon{0009-0001-0394-650X}}
\email{rahmatovbekzod@samdu.uz}
\affiliation{New Uzbekistan University, Movarounnahr str. 1, Tashkent 100000, Uzbekistan}

\author{Shokhzod~Jumaniyozov\orcidicon{0009-0009-6254-5608}}
\email{sh.jumaniyozov@newuu.uz}
\affiliation{Kimyo International University in Tashkent, Shota Rustaveli Street 156, Tashkent 100121, Uzbekistan}

\author{Dilafruz Xujanova \orcidicon{0009-0006-0709-9359}}
\email{xujanova1980@gmail.com}  
\affiliation{University of Economics and Pedagogy, 13 Islam Karimov St., Karshi, 180100, Uzbekistan}

\author{Farhod~Nuraliev \orcidicon{0009-0009-1711-2088}}
\email{nuraliyevf@mail.ru}
\affiliation{Tashkent International University, Little Ring Road 7, Tashkent 100115, Uzbekistan}
\date{\today}

\begin{abstract}
We study a static, spherically symmetric black hole solution surrounded by a dust field within quantum fluctuation modified gravity, in which the quantum-fluctuation parameter $\alpha$ and the dust-field strength parameter $K$. The horizon structure reveals the two parameters acting in opposite directions: strengthening $K$ draws the inner and outer horizons together until they merge into a single extremal horizon, while a larger $\alpha$ restores the Schwarzschild limit; the same opposing trend extends to the effective potential and the characteristic circular-orbit radii -- the innermost stable circular orbit, the marginally bound orbit, and the marginal circular orbit -- which shift inward with $K$ and outward with $\alpha$, and to the Keplerian and radial epicyclic frequencies of massive test particles. From these frequencies we build three twin-peak quasi-periodic oscillation models -- relativistic precession, warped-disk, and epicyclic-resonance -- along with their $3{:}2$, $4{:}3$, and $5{:}4$ resonance radii, all following the same $K$-$\alpha$ trend. Testing all three models against twin-peak QPO data from two stellar-mass (GRO~J1655$-$40, XTE~J1550$-$564) and two intermediate-mass (M82~X$-$1, NGC~1313~X$-$1) sources through a Markov Chain Monte Carlo analysis places best-fit constraints, at the $2\sigma$ level, of $K\in(0,0.5)$ and $\alpha\in(0.3,0.6)$ across all sources and models.
\vspace{6pt}

\textbf{Keywords:}{Quantum fluctuation modified gravity, Dust field, QPO}
\end{abstract}
\maketitle
\section{Introduction}
The observed late-time acceleration of cosmic expansion \cite{1998AJ....116.1009R,1999ApJ...517..565P} has motivated a wide range of modified gravity proposals, including $f(R)$ gravity \cite{2004PhRvD..70d3528C}, and $f(R,T)$ gravity \cite{2011PhRvD..84b4020H}. A more recent addition to this landscape, quantum fluctuation modified gravity (QFMG), originates from Heisenberg's non--perturbative quantization of the gravitational field \citep{2014EPJC...74.2743D,Dzhunushaliev:2015hoa}: splitting the metric operator into a classical background and a quantum-fluctuation piece generates, at the classical level, a non-minimal coupling between geometry and matter \citep{2016PDU....13...87Y}. This construction has since been developed in several directions \citep{2016PDU....13...87Y,2016EPJC...76..420L,Bernardo:2021ynf,Chen:2021oal,Lima:2023qdv}, including black hole solutions \citep{2021PDU....3100756Y}, Friedmann-Lema\^itre-Robertson-Walker cosmology \citep{2016PDU....13...87Y,2016EPJC...76..420L,Bernardo:2021ynf,Haghani:2021iqe,Chen:2021oal}, and, most recently, baryogenesis \citep{Yang:2024gnf}.

Black hole solutions offer a natural testing ground for such modified theories, since any matter-geometry coupling is expected to leave its clearest imprint in the high-density, strong-field regime rather than at cosmological scales; the influence of this coupling on black hole structure within QFMG was examined in \citep{2021PDU....3100756Y}. Here we seek a black hole solution of the QFMG field equations sourced by the anisotropic field introduced by Kiselev \citep{2003CQGra..20.1187K}, who related the field's energy density and pressure through a barotropic equation of state obtained by averaging the anisotropic stress-energy tensor over angles; specific choices of the equation-of-state parameter within this formulation can reproduce an accelerating cosmological background \citep{2003CQGra..20.1187K}. Building on this framework, numerous Kiselev-type black hole solutions have since been proposed and investigated across a range of gravitational settings \citep{Ghosh:2015ovj,Xu:2016jod,Toshmatov:2015npp,Singh:2023zmy,Sood:2024ufi,Bezerra:2022srj,Gogoi:2024ypn,Sheikhahmadi:2023jpb}.

The general black hole solution in anisotropic field within QFMG, valid for an arbitrary equation-of-state parameter $\omega$, was recently obtained by \citep{2026EPJP..141..135H}. In the present work we specialize this solution to the pressureless dust case, $\omega=0$, which is of particular astrophysical interest as a simple model for the cold, non-relativistic matter -- such as interstellar dust or a dark-matter component -- that may surround an accreting compact object.

Beyond these geometric properties, twin-peak quasi-periodic oscillations (QPOs) observed in the X-ray flux of accreting black holes provide a direct observational probe of the strong-field orbital structure \citep{2006ARA&A..44...49R}, and have been used extensively to test deviations from the Schwarzschild and Kerr geometries in a variety of modified-gravity and exotic black hole backgrounds \citep{2022EPJC...82.1110R,2024ChPhC..48e5104R,2023EPJC...83..572R}.

The paper is organized as follows. Sec.~\ref{sec2} derives the QFMG field equations and the dust-field black hole solution and analyse the black hole horizon structure. In sec.~\ref{sec3} we examine the resulting  and circular geodesic motion. Sec.~\ref{sec.4} derives the fundamental orbital frequencies. Sec.~\ref{sec:qpo} constructs the RP, WD, and ER4 QPO models together with their resonance radii. Sec.~\ref{section6} presents the MCMC analysis and its results. In Sec.~\ref{sec.7} concludes.

\section{Black Hole Solution in QFMG surrounded by anisotropic field \label{sec2}}
In this section, we briefly review a black hole solution in QFMG \cite{2026EPJP..141..135H}. In Heisenberg's nonperturbative quantization scheme, the metric operator is split into a classical background $g_{\mu\nu}$ and a quantum fluctuation part $\delta\widehat{g}_{\mu\nu}$, so that $\hat{g}_{\mu\nu}=g_{\mu\nu}+\delta\widehat{g}_{\mu\nu}$ \cite{2014EPJC...74.2743D}. As shown in Ref. \cite{2017PhRvD..95l3525W} the fluctuation $\delta\hat{g}^{\mu\nu}$ decomposes into a physical part and a gauge part, the latter removable through a suitable gauge transformation, while $\langle\delta\hat{g}_{\mu\nu}\rangle$ remains generally nonzero for any non-vacuum quantum state. Expanding the quantum Einstein–Hilbert Lagrangian to first order in this fluctuation yields
\begin{eqnarray}
\label{lag0}
L_{\hat{g}}(\hat{g})=L_{\hat{g}}(g+\delta\hat{g})\approx L_{g}(g)+\frac{\delta L_{g}}{\delta g^{\mu\nu}}\delta \widehat{g}^{\mu\nu}.
\end{eqnarray}
Here $c=\hbar=1$ and $k^2=8\pi G$. Because $\left\langle \frac{\delta L_{g}}{\delta g^{\mu\nu}}\delta \widehat{g}^{\mu\nu}\right\rangle=\frac{\delta L_{g}}{\delta g^{\mu\nu}}\langle\delta \widehat{g}^{\mu\nu}\rangle=\sqrt{-g}G_{\mu\nu}\langle\delta \widehat{g}^{\mu\nu}\rangle$ \cite{2014EPJC...74.2743D}, the expectation value of Lagrangian (\ref{lag0}) is written as
\begin{eqnarray}
\label{lag00}
\langle L_{\hat{g}}\rangle\approx \frac{1}{2k^2}\sqrt{-g} \Big[R+G_{\mu\nu}\langle\delta \hat{g}^{\mu\nu}\rangle \Big],
\end{eqnarray}
The matter Lagrange density $L_{\rm m}^{\hat{g}}$ can likewise be expanded to first order in the fluctuation
\begin{eqnarray}
\label{lagm}
L^{\hat{g}}_{\rm m}(g+\delta\hat{g})\approx \sqrt{-g}L_{\rm m}(g)+\frac{\delta \sqrt{-g}L_{\rm m}}{\delta g^{\mu\nu}}\delta \widehat{g}^{\mu\nu}.
\end{eqnarray}
Taking the expectation value of Eq.~(\ref{lagm}) then gives \cite{2014EPJC...74.2743D}
\begin{eqnarray}
\label{lagm0}
\langle L^{\hat{g}}_{\rm m}(g+\delta\hat{g}) \rangle \approx \sqrt{-g}\bigg[L_{\rm m}-\frac{1}{2}T_{\mu\nu}\langle\delta \hat{g}^{\mu\nu}\rangle\bigg],
\end{eqnarray}
where $T_{\mu\nu}=-2\delta(\sqrt{-g}L_{\rm m})/(\sqrt{-g}\delta g^{\mu\nu})$ denotes the stress-energy tensor, so that the modified Lagrangian density takes the form
\begin{eqnarray}
\label{lag}
L=\frac{1}{2k^2}\sqrt{-g} \bigg[R+G_{\mu\nu}\langle\delta \hat{g}^{\mu\nu}\rangle \bigg]+\sqrt{-g}\bigg[L_{\rm m}-\frac{1}{2}T_{\mu\nu}\langle\delta \hat{g}^{\mu\nu}\rangle\bigg].
\end{eqnarray}
For a simple case, $\langle\delta \hat{g}^{\mu\nu}\rangle=\alpha g^{\mu\nu}$ with $\alpha$ is quantum--fluctuation parameter and it indicates the magnitude of quantum fluctuations, which fulfills the two conditions suggested in \cite{2017PhRvD..95l3525W}, $\nabla_\alpha\langle\delta \hat{g}^{\mu\nu}\rangle=0$ and $\langle\delta \hat{g}^{\mu\nu}\rangle=\langle\delta \hat{g}^{\nu\mu}\rangle$, the Lagrangian density (\ref{lag}) becomes \cite{2016PDU....13...87Y},
\begin{eqnarray}
\label{act}
L=L_{\rm mg}+L_{\rm mm}=\frac{1}{2k^2}\sqrt{-g}(1-\alpha)R+\sqrt{-g}\bigg[L_{\rm m}-\frac{1}{2}\alpha T\bigg],
\end{eqnarray}
Here $T=g_{\mu\nu}T^{\mu\nu}$ denotes the trace of the stress-energy tensor, and we work throughout in the regime $|\alpha|<1$, for which the quantum-fluctuation contribution stays subdominant to the classical metric. Imposing $\delta g_{\mu\nu}=0$ at the boundary and varying the Lagrangian density~(\ref{lag}) with respect to $g^{\mu\nu}$ then leads to the gravitational field equations \cite{2016PDU....13...87Y}
\begin{eqnarray}
\label{mot}
G_{\mu\nu}\equiv R_{\mu\nu}-\frac{1}{2}g_{\mu\nu}R=\frac{2k^2}{1-\alpha}\left[\frac{1}{2}(1+\alpha)T_{\mu\nu}-\frac{1}{4}\alpha g_{\mu\nu}T+\frac{1}{2}\alpha\theta_{\mu\nu}\right],
\end{eqnarray}
with $\theta_{\mu\nu}\equiv g^{\alpha\beta}\delta T_{\alpha\beta}/\delta g^{\mu\nu}$ and $\theta\equiv g_{\mu\nu}\theta^{\mu\nu}$. Using $R=-k^2[T+\alpha\theta/(1-\alpha)]$, Eq.~(\ref{mot}) can equivalently be cast in the Ricci-tensor form \cite{2016PDU....13...87Y}
\begin{eqnarray}
\label{mot1}
R_{\mu\nu}=\frac{2k^2}{1-\alpha}\left[\frac{1}{2}(1+\alpha)T_{\mu\nu}-\frac{1}{4} g_{\mu\nu}T+\frac{1}{2}\alpha\theta_{\mu\nu}-\frac{1}{4}\alpha g_{\mu\nu}\theta\right].
\end{eqnarray}
The Lagrangian density~(\ref{lag}) offers a possible microscopic origin for matter-creation processes within $f(R,T)$ or $f(R,L_{\rm m})$ gravity \citep{2016EPJC...76..420L,2021PDU....3100756Y}, potentially clarifying the physical mechanism by which particle production proceeds through matter-geometry coupling. 

For a spherically symmetric spacetime, the line element can be expressed as:
\begin{eqnarray}
\label{lineelement}
ds^{2}=-f(r)dt^{2}+g(r)dr^{2}+r^{2}\Big(d\theta^{2}+\sin^{2}\theta d\phi^{2}\Big),
\end{eqnarray}
with $f(r)$ and $g(r)$ analytic functions of the radial coordinate $r$. For the Kiselev anisotropic field, the energy-momentum tensor components take the form \citep{2003CQGra..20.1187K}
\begin{eqnarray}
	&& T^{t}_{t}=T^{r}_{r}=\rho(r), \label{T12} \\
	&& T^{\theta}_{\theta}=T^{\phi}_{\phi}=-\frac{1}{2}\rho(3\omega+1),    \label{T34}
\end{eqnarray}
Here $\rho$ denotes the energy density and $\omega=p/\rho$ the equation-of-state parameter, obtained by isotropically averaging over the angular directions in Eqs.~(\ref{T12}) and~(\ref{T34}) \citep{2003CQGra..20.1187K}, with $p$ the pressure. The corresponding energy-momentum tensor of the anisotropic Kiselev field then reads
\begin{eqnarray}
	\label{TTT}
	T^{\mu}_{\nu}={\rm{diag}}(\rho,-p_{r},-p_{t},-p_{t}),
\end{eqnarray}
Here $p_t$, $\rho$, and $p_r$ denote the transverse pressure, energy density, and radial pressure of the field, respectively. Comparison with Eqs.~\eqref{T12} and~\eqref{TTT} fixes $p_r=-\rho$ and $p_t=\frac{1}{2}\rho(3\omega+1)$. The corresponding matter Lagrangian density, $L_m=-(1/3)(p_r+2p_t)$, yields $\theta_{\mu\nu}=-2T_{\mu\nu}-\frac{1}{3}(p_r+2p_t)g_{\mu\nu}$; within the Kiselev field, consistency of these relations requires additivity and linearity among the metric components. 

Substituting this result into the Einstein tensor, we derive

\begin{eqnarray}
	&&\frac{1}{r}\frac{df(r)}{dr}-\frac{f(r)}{r^{2}}+\frac{1}{r^{2}}=\frac{2k^2}{1-\alpha}\left[\frac{1}{2}\rho(1+\alpha)-\frac{1}{4}\alpha\rho(1-3\omega)-\frac{1}{2}\alpha\rho(2+\omega)\right]                          \label{GGH}  \\
	&&-\frac{1}{2}\frac{d^{2}f(r)}{dr^{2}}-\frac{1}{r}\frac{df(r)}{dr}=\frac{2k^2}{1-\alpha}\left[-\frac{1}{4}\rho(1+\alpha)(3\omega+1)-\frac{1}{4}\alpha\rho (1-3\omega)+\frac{1}{2}\alpha\rho(2\omega+1)\right]          \label{GGF}
\end{eqnarray}

Combining equation with Eqs. (\ref{GGH}) and (\ref{GGF}), gives
\begin{eqnarray}
\label{integrate1} -\frac{1}{r}\frac{df(r)}{dr}-\frac{f(r)}{r^{2}}+\frac{1}{r^{2}}=
\frac{2-3\alpha+\omega\alpha}{4\omega\alpha-3\omega-1}\left[-\frac{1}{2}\frac{d^{2}f(r)}{dr^{2}}-\frac{1}{r}\frac{df(r)}{dr}\right].
\end{eqnarray}

Eq. (\ref{integrate1}) can simply as
\begin{eqnarray}
	\label{integrate2}
	\frac{1}{r^{2}}\left[\frac{d}{dr}\left(rf(r)\right)-1\right]=
-\frac{1}{2r}\frac{2-3\alpha+\omega\alpha}{1-4\omega\alpha+3\omega}\frac{d}{dr}\left[\frac{d}{dr}\left(rf(r)\right)-1\right].
\end{eqnarray}
After integrating, we get
\begin{eqnarray}
	\label{integrate3}
	\frac{d}{dr}\left[rf(r)\right]=1+C_1 r^{\frac{-2(1+3\omega-4\omega\alpha)}{2-3\alpha+\omega\alpha}},
\end{eqnarray}
where $C_1$ is an integration constant. Integrating again

\begin{eqnarray}\label{eqB1}
rf(r)=r+\frac{C_1}{1-\frac{2(1+3\omega-4\omega\alpha)}{2-3\alpha+\omega\alpha}}r^{1-\frac{2(1+3\omega-4\omega\alpha)}{2-3\alpha+\omega\alpha}}+C_2, \end{eqnarray} 
to simplify above Eq.(\ref{eqB1}) and it reduces to the Schwarzschild black hole, fixing $C_2=-2M$ with $M$ the black hole mass:

\begin{eqnarray}\label{eqB}
f(r)=1-\frac{2M}{r}+C_1\frac{2-3\alpha+\omega}{3(3\omega\alpha-\alpha-2\omega)}r^{-\frac{2(1+3\omega-4\omega\alpha)}{2-3\alpha+\omega\alpha}}. \end{eqnarray} 

If $\alpha\to0$ limiting, Eq.(\ref{eqB}) reduces to Kiselev black hole solution \cite{2003CQGra..20.1187K}: 
\begin{eqnarray}\label{eqBKiselev}
f(r)=1-\frac{2M}{r}-\frac{N}{r^{3\omega+1}}
\end{eqnarray}
with $N\equiv -C_1/(3\omega)$ the normalization constant of the surrounding field.

\subsection{Horizon structure of black hole surrounded by dust field in QFMG}
For this work, we consider the $\omega=0$ (dust field) case \cite{2003CQGra..20.1187K,1972PhRvD...6.3357R}, and Eq.~(\ref{eqB}) reduces to
\begin{eqnarray}\label{eqBdust}
f(r)=1-\frac{2M}{r}-K\frac{3\alpha-2}{3\alpha}r^{\frac{2}{3\alpha-2}}.
\end{eqnarray}

where $C_1=-K$, $K$ characterizes the strength of the surrounding dust field. It should be noted that, the apparent divergence of in Eq. (\ref{integrate3}) the coefficients as $\alpha\to0$ and $\omega\to0$ are a removable artifact of the integration exponent approaching the resonant value $\frac{-2(1+3\omega-4\omega\alpha)}{2-3\alpha+\omega\alpha}\to-1$, at which the power-law antiderivative is replaced by a logarithm; absorbing the divergent piece into the second integration constant, the $\alpha\to0\,\& \, \omega\to0$ limits are smooth and yields
\begin{eqnarray}\label{eqBlog}
\nonumber f(r)=1-\frac{2M}{r}+C_1\frac{\ln r}{r}.
\end{eqnarray}
\begin{figure}
\centering
\includegraphics[width=0.45\linewidth]{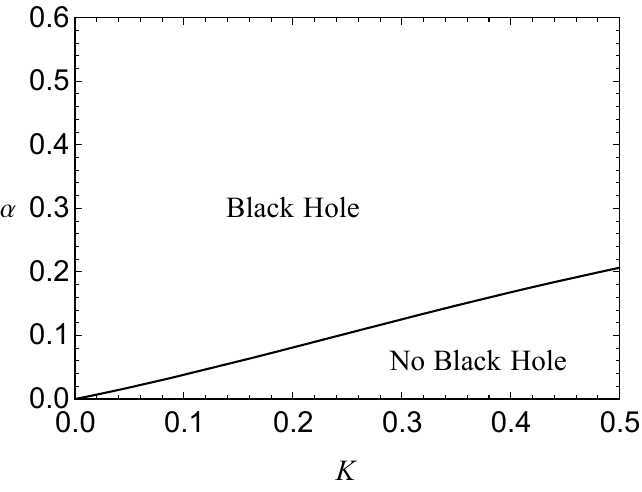}
    \caption{Black hole and no black hole regions in the $(K,\alpha)$ plane for the dust-field QFMG black hole. Here $M=1$. }
    \label{fig:bh-nbh}
\end{figure}

To determine whether a black hole horizon exists for given values of $K$ and $\alpha$, we solve the horizon conditions $f(r)=0$ and $\partial f(r)/\partial r=0$ simultaneously using Eq.~(\ref{eqBdust}). These equations yield the critical relation between $K$ and $\alpha$. The resulting parameter space, indicating whether a horizon exists or not, is presented in Fig.~\ref{fig:bh-nbh}. We adopt the intervals $K\in(0,0.5)$ and $\alpha\in(0.2,0.6)$ for our subsequent analysis.
\begin{figure}
    \centering
    \includegraphics[width=0.48\linewidth]{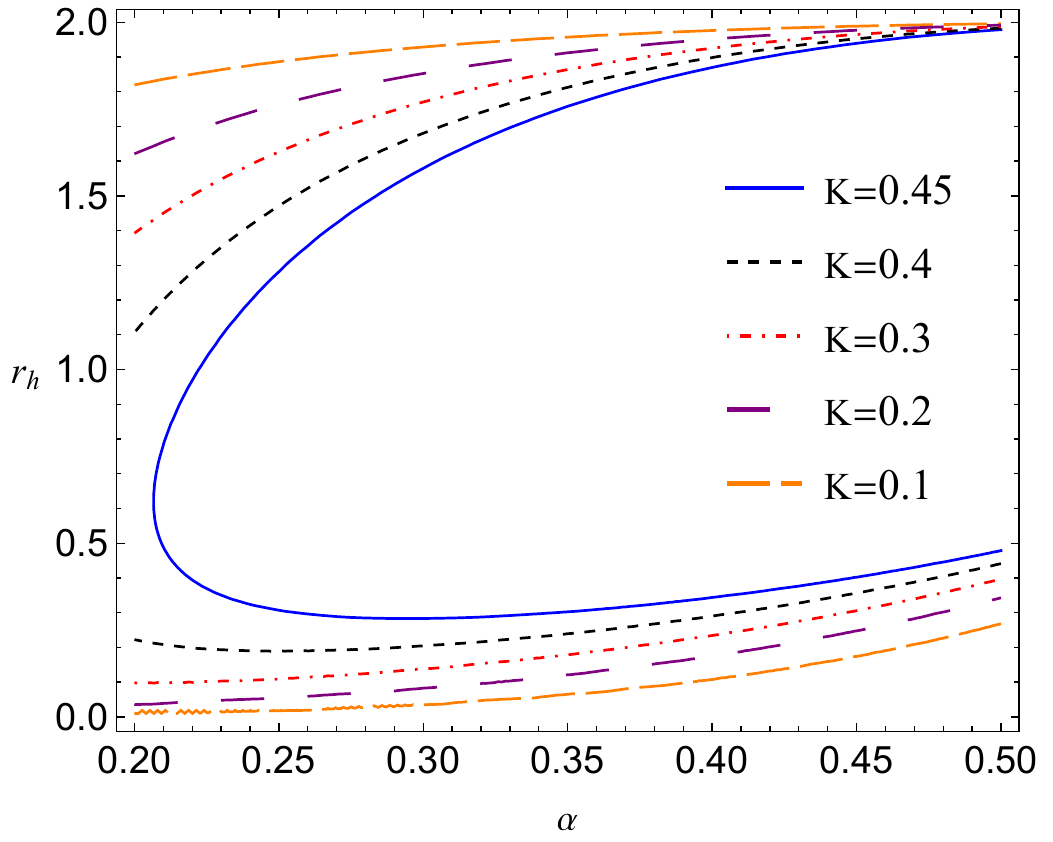}
   \caption{Horizon radii $r_h$ as a function of the 
quantum-fluctuation parameter $\alpha$ for selected values 
of the dust-field strength $K$ for the dust-field QFMG 
black hole.}
    \label{fig:fr-alpha-k}
\end{figure}
Figure~\ref{fig:fr-alpha-k} shows the horizon radii $r_h$ as functions of $\alpha$ for several values of $K$. At fixed $K$, increasing $\alpha$ raises both the inner/Cauchy $(r_-)$ and outer/event $(r_+)$ horizons, while at fixed $\alpha$, increasing $K$ raises the $(r_-)$ and lowers the $(r_+)$; the two horizons therefore approach each other as $K$ grows, and for sufficiently strong dust field they merge into a single extremal horizon -- the merger visible for the blue curve traces the critical boundary already identified in Fig.~\ref{fig:bh-nbh}. Over the studied range, the inner horizon remains sensitive to $K$ throughout, whereas the outer horizon's dependence on $K$ weakens as $\alpha$ increases and becomes negligible as $\alpha\to0.5$, where all curves converge to the Schwarzschild value $r_+=r_{\text{Schw.}}=2M$.

\section{Test particle motion around the black hole surrounded by a dust field in QFMG \label{sec3}} 

We examine timelike geodesics in the dust-field QFMG spacetime. The conserved specific energy and angular momentum associated with the Killing vectors $\partial_t$ and $\partial_\phi$ are given by 
\begin{eqnarray} 
E=f(r)\dot{t}, \qquad L=r^{2}\dot{\phi}. 
\end{eqnarray} 
Restricting the motion to the equatorial plane $\theta=\pi/2$ and applying the normalization condition $g_{\mu\nu}\dot{x}^{\mu}\dot{x}^{\nu}=-1$, the radial equation takes the effective-potential form 
\begin{eqnarray} 
\dot{r}^{2}+V_{\rm eff}(r)=E^{2}, 
\end{eqnarray} 
where 
\begin{eqnarray} 
V_{\rm eff}(r)=f(r)\left(1+\frac{L^{2}}{r^{2}}\right). 
\end{eqnarray} 
For the dust-field QFMG metric~\eqref{eqBdust}, the effective potential becomes \begin{eqnarray} V_{\rm eff}(r)=\left[1-\frac{2M}{r}-\frac{(3\alpha-2)K}{3\alpha}\,r^{\frac{2}{3\alpha-2}}\right]\left(1+\frac{L^{2}}{r^{2}}\right). \end{eqnarray}  
\begin{figure*}
    \centering
\includegraphics[width=0.49\linewidth]{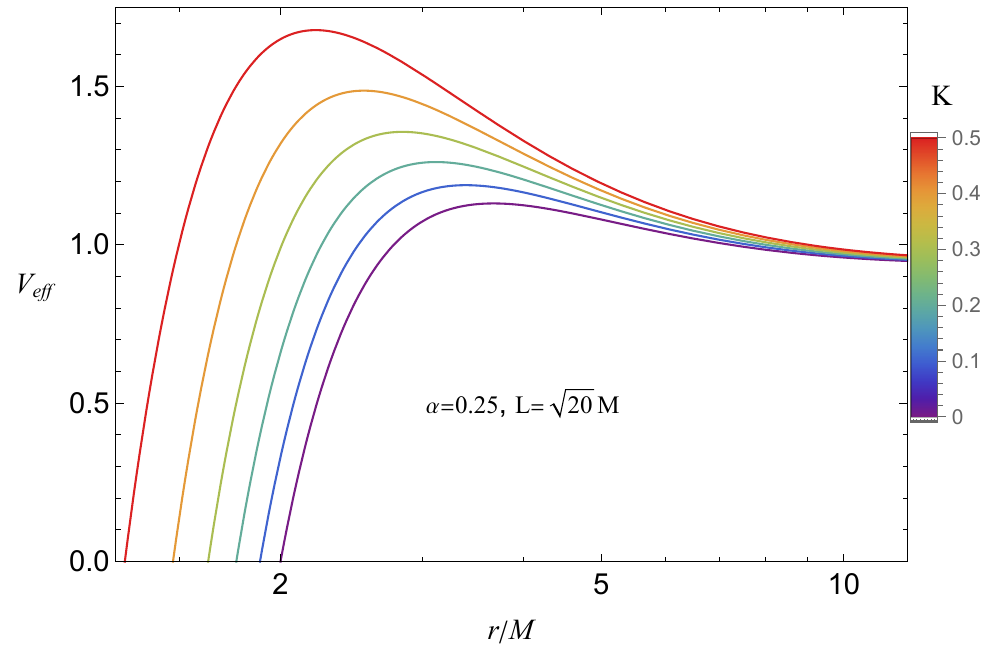}
\includegraphics[width=0.49\linewidth]{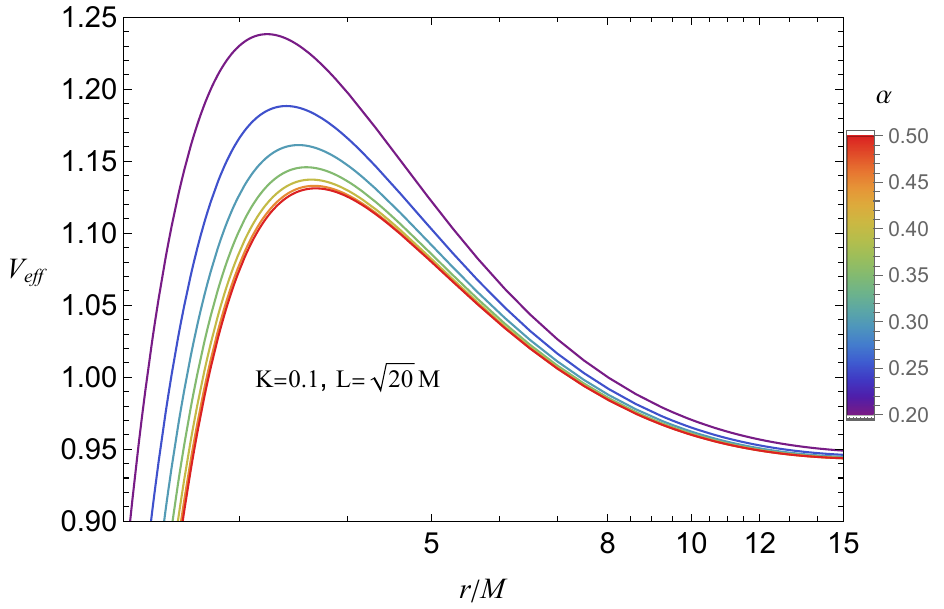}
\caption{Effective potential $V_{\rm eff}$ as a function of $r/M$ for selected values of $K$ (left panel) and $\alpha$ (right panel) for the dust-field QFMG black hole.}
   \label{V.eff}
\end{figure*}

Fig.~\ref{V.eff} displays the effective potential $V_{\rm eff}$ for representative values of $K$ (left panel, $\alpha=0.25$) and $\alpha$ (right panel, $K=0.1$). Increasing $K$ raises the potential barrier and shifts its maximum inward, showing that a stronger dust field strengthens the gravitational effect near the black hole and pulls the associated unstable circular orbit closer to the horizon. Increasing $\alpha$ instead lowers the barrier and pushes its maximum outward, from $r\sim3.1M$ to $r\sim3.7M$ over the range considered, so that stronger quantum-fluctuation corrections weaken this effect and displace the unstable orbit farther out. At large $r$, both panels converge onto a single curve irrespective of $K$ and $\alpha$, reflecting the asymptotic recovery of the Schwarzschild geometry, where the dust-field and quantum-fluctuation modifications become dynamically irrelevant.

\subsection{Circular motion} Circular orbits are characterized by a constant radial coordinate and are determined by the conditions 
\begin{eqnarray} 
V_{\rm eff}(r)=E^{2}, \qquad \frac{dV_{\rm eff}}{dr}=0.
\end{eqnarray} 
For a static and spherically symmetric spacetime, these conditions yield 
\begin{eqnarray} &&\mathcal{L}^{2}_{c}=\frac{r^{3}f'(r)}{2f(r)-rf'(r)}=\frac{r^{2}\left(\frac {K}{\alpha} r^{\frac{2}{3\alpha-2}+1}-3M\right)}{3\left(-\frac {K}{\alpha}r^{\frac{2}{3\alpha-2}+1}+ Kr^{\frac{2}{3\alpha-2}+1}+3 M- r\right)} \\&&\mathcal{E}^{2}_{c}=\frac{2f^{2}(r)}{2f(r)-rf'(r)}=\frac{\left((3\alpha-2)\frac {K}{\alpha}r^{\frac{3\alpha}{3\alpha-2}}+6 M-3 r\right)^{2}}{9 r\left((r-3M)-(\alpha-1)\frac {K}{\alpha}r^{\frac{3\alpha}{3\alpha-2}}\right)}. 
\label{eq:Ec_Lc_general}
\end{eqnarray} 
where $\mathcal{E}_c$ and $\mathcal{L}_c$ are the specific energy and angular momentum of circular motion, respectively. Substituting the dust-field QFMG metric~\eqref{eqBdust}, one obtains 

In the Schwarzschild limit $K=0$, these expressions reduce to the standard results $\mathcal{L}^{2}_{c}=Mr^{2}/(r-3M)$ and $\mathcal{E}^{2}_{c}=(r-2M)^{2}/[r(r-3M)]$.

\begin{figure*}
\centering
\includegraphics[width=0.49\linewidth]{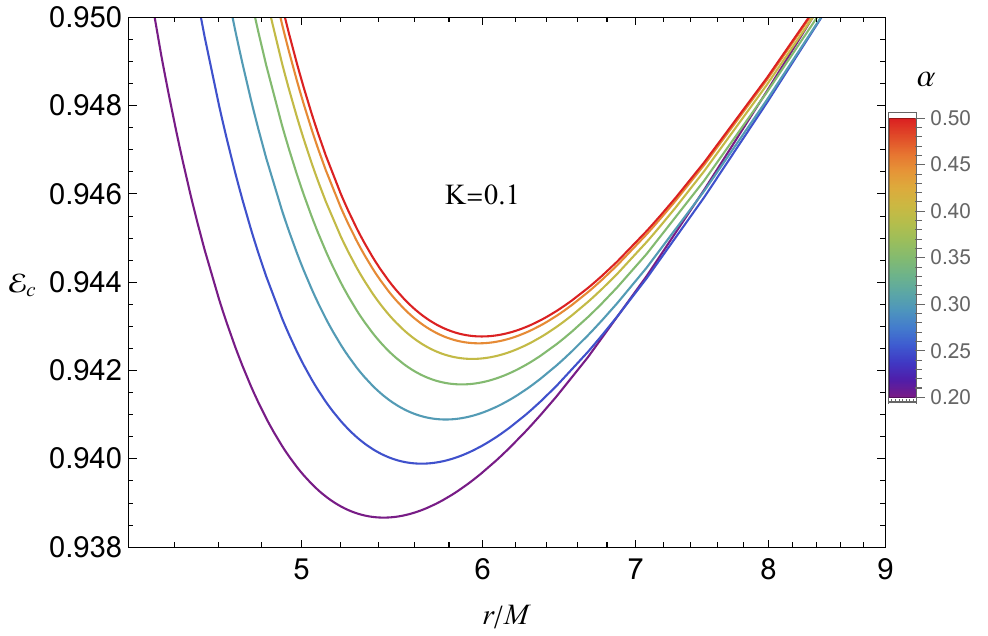}
\includegraphics[width=0.49\linewidth]{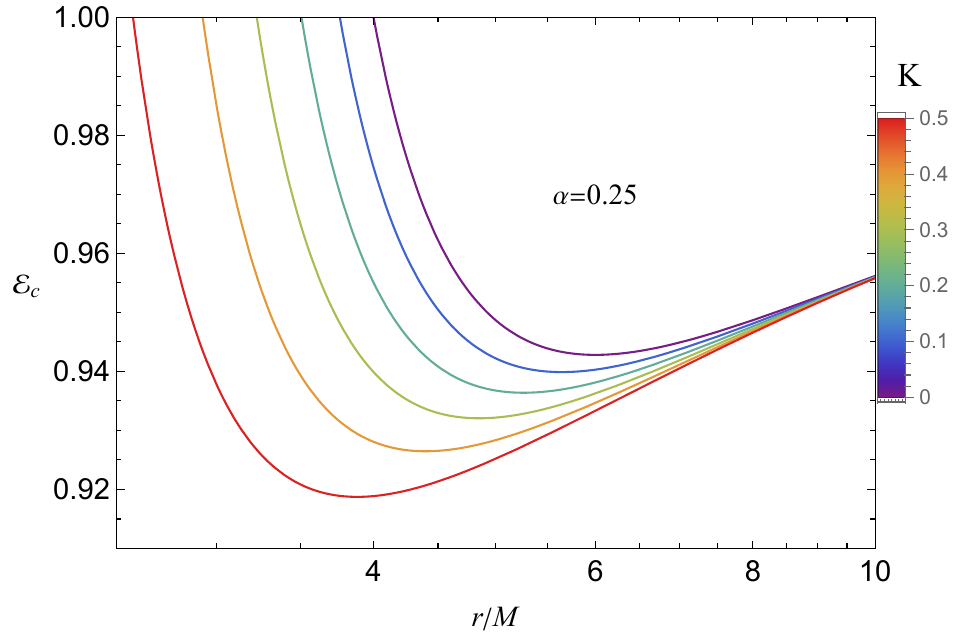}
\includegraphics[width=0.49\linewidth]{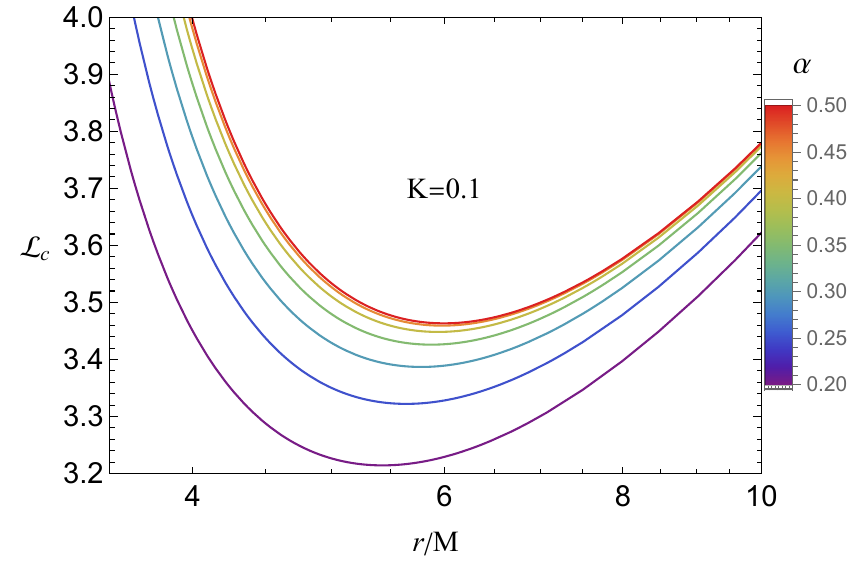}
\includegraphics[width=0.49\linewidth]{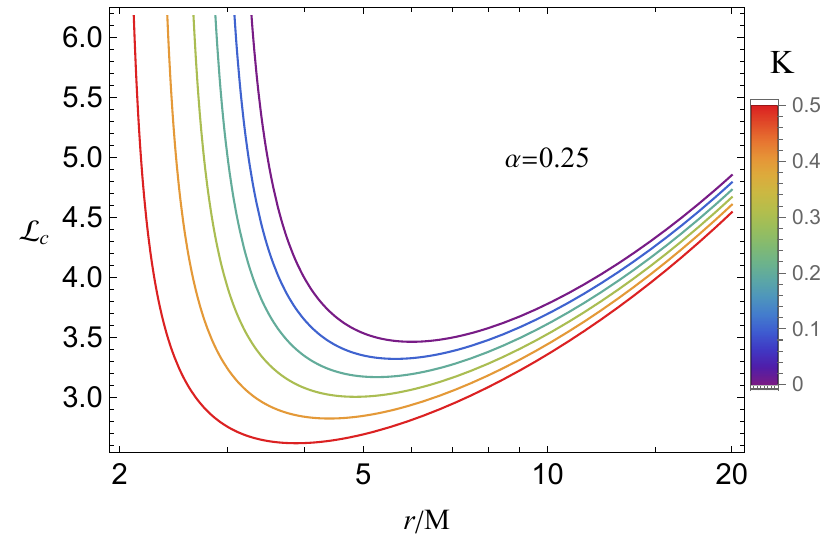}
\caption{Specific energy $\mathcal{E}_c$ (upper panels) and angular momentum $\mathcal{L}_c$ (lower panels) for circular orbits in the dust-field QFMG black hole.}
\label{Ec vs Lc}
\end{figure*}

Figure~\ref{Ec vs Lc} shows the specific energy $\mathcal{E}_c$ (upper panels) and angular momentum $\mathcal{L}_c$ (lower panels) of circular orbits in the dust-field QFMG spacetime. Near the black hole, both $\mathcal{E}_c$ and $\mathcal{L}_c$ rise sharply, pass through a minimum, and increase again at large $r$; this minimum marks the location of the innermost stable circular orbit (ISCO), discussed in detail in the following subsection. In the upper-left panel, increasing $\alpha$ raises the minimum of $\mathcal{E}_c$ and shifts it outward, so that a stronger QF effect pushes the ISCO to larger radii and increases the binding energy of the innermost bound orbit. In the upper-right panel, increasing $K$ instead lowers the minimum and shifts it inward, so that a stronger dust field pulls the ISCO closer to the black hole and reduces this binding energy. For $K=0$, the Schwarzschild result is recovered for all $\alpha$. The lower panels display the same pattern for $\mathcal{L}_c$: increasing $\alpha$ at fixed $K$ shifts its minimum outward, while increasing $K$ at fixed $\alpha$ shifts it inward, consistent with the outward and inward displacement of the ISCO noted above. At large $r$, as seen for the effective potential in Fig.~\ref{V.eff}, all curves converge onto a single branch, reflecting the recovery of the Schwarzschild geometry far from the black hole.

\subsection{ISCO} 
ISCO marks the inner boundary of the region where stable circular motion is possible: for $r>r_{\rm ISCO}$ a test particle can orbit stably, whereas for $r<r_{\rm ISCO}$ any small perturbation causes it to plunge into the black hole. It therefore sets the inner edge of a geometrically thin accretion disk and provides the length scale against which the QPO resonance radii of Sec.~\ref{sec:qpo} are measured, making its dependence on $K$ and $\alpha$ directly relevant for observational tests of the model.

The stability of a circular orbit against radial perturbations is governed by the sign of $d^{2}V_{\rm eff}/dr^{2}$ evaluated at that orbit: stable circular motion requires
\begin{eqnarray}
\frac{d^{2}V_{\rm eff}}{dr^{2}}\bigg|_{r=r_{0}}>0,\,\,\, \& \,\,\, \frac{d^{2}V_{\rm eff}}{dr^{2}}\bigg|_{r=r_{\rm ISCO}}=0.
\end{eqnarray}
For the dust-field QFMG metric, the ISCO radius follows from the root of
\begin{eqnarray}
K\left[2(\alpha-1)(3\alpha-2)Kr^{\frac{6\alpha}{3\alpha-2}}-3\alpha[\alpha(3\alpha-16)+8]Mr^{\frac{3\alpha}{3\alpha-2}}-2\alpha(3\alpha-1)r^{\frac{2(3\alpha-1)}{3\alpha-2}}\right]+3\alpha^{2}(3\alpha-2)M(r-6M)=0. \label{isco}
\end{eqnarray}
Setting $K=0$ removes the entire bracketed term, so Eq.~(\ref{isco}) reduces to the Schwarzschild value $r=6M$.

The top panels of Fig.~\ref{f5} shows how $r_{\rm ISCO}$ depends on $\alpha$ and $K$. In the top-left panel, the Schwarzschild curve $K=0$ stays flat at $r_{\rm ISCO}=6M$ for all $\alpha$, while every curve with $K\neq0$ rises with $\alpha$ and reaches $6M$ only as $\alpha\to0.5$; at fixed $\alpha$, larger $K$ systematically pulls $r_{\rm ISCO}$ to smaller radii, so a stronger dust field draws the innermost stable orbit inward. The top-right panel shows the same interplay viewed differently: at fixed $\alpha$, increasing $K$ steadily lowers $r_{\rm ISCO}$, with a steep decline for small $\alpha$ -- where even a modest dust field produces a large inward shift -- and an increasingly shallow one as $\alpha\to0.5$, where $r_{\rm ISCO}$ remains close to $6M$ even at $K=0.5$. Together, the two panels show that $\alpha$ and $K$ act in opposite directions: the quantum-fluctuation parameter pushes the ISCO outward while the dust field pulls it inward, so $r_{\rm ISCO}$ provides insight into this interplay in the strong-field orbital structure.
For $K=0$, this reduces to the Schwarzschild result $r_{\rm ISCO}=6M$.

\subsection{Marginally bound orbit}

The marginally bound orbit (MBO) identifies the circular geodesic at which the specific energy equals unity, $\mathcal{E}_c=1$; it separates orbits that remain gravitationally bound, capable of settling onto an eccentric trajectory once perturbed, from those with enough energy to escape to infinity, and thus sets the relevant scale for capture and tidal-disruption processes near the black hole.

Combined with the circular-orbit conditions, this criterion reduces at the metric level to
\begin{equation}
2f^{2}(r)-2f(r)+rf'(r)=0,
\end{equation}
which for the dust-field QFMG metric takes the explicit form
\begin{equation}
(3\alpha-2)^2K^2 r^{\frac{6\alpha}{3\alpha-2}}
+3\alpha K r^{\frac{3\alpha}{3\alpha-2}}
\left[4M(3\alpha-2)-(3\alpha-1)r\right]
+9\alpha^2 M(r-4M)=0.
\label{eq:mbo_qfmg_simplified}
\end{equation}

Setting $K=0$ collapses this to $9\alpha^2M(r-4M)=0$, recovering the Schwarzschild value $r_{\rm MBO}=4M$.

\begin{figure*}
    \centering    
    \includegraphics[width=0.49\linewidth]{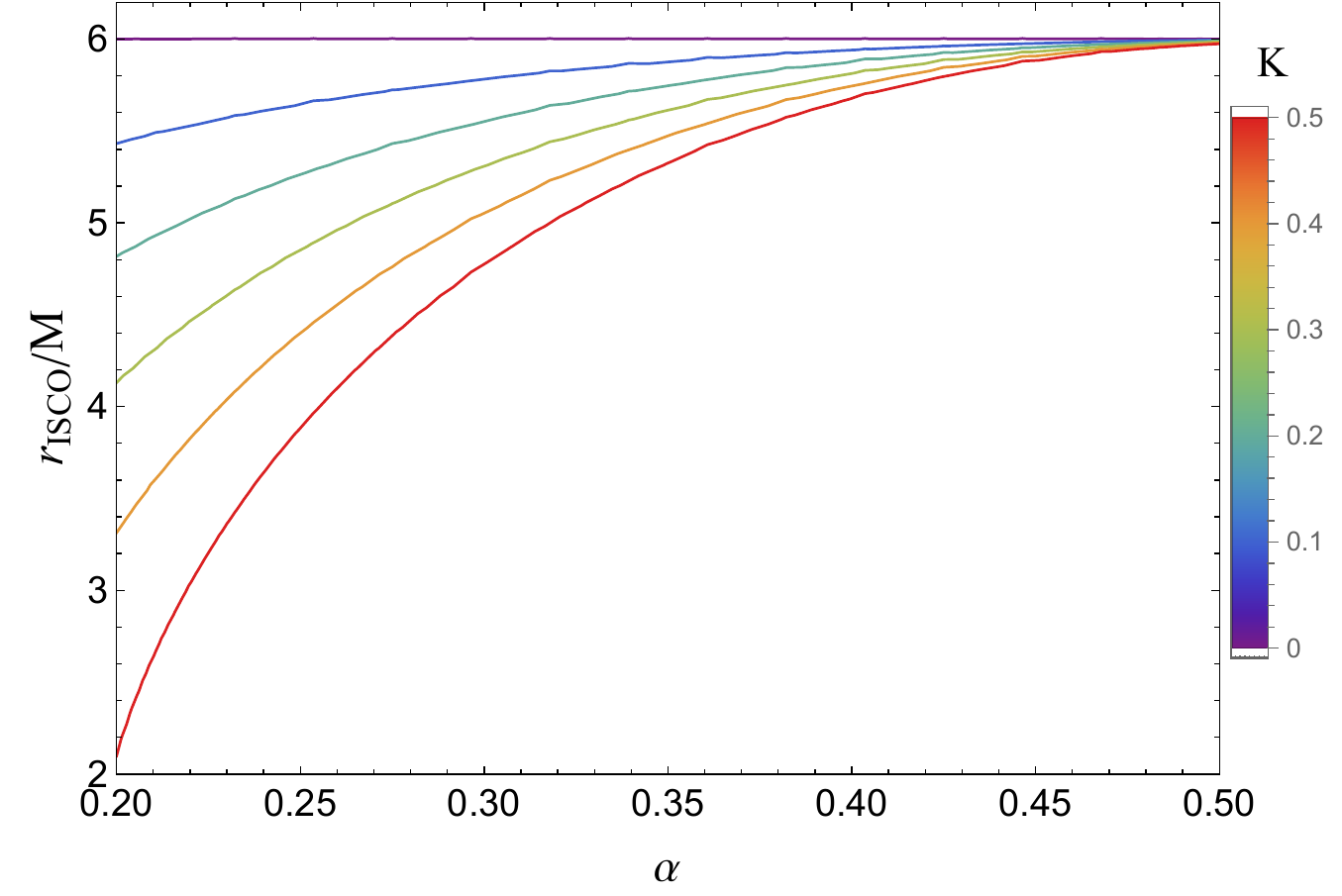}
    \includegraphics[width=0.49\linewidth]{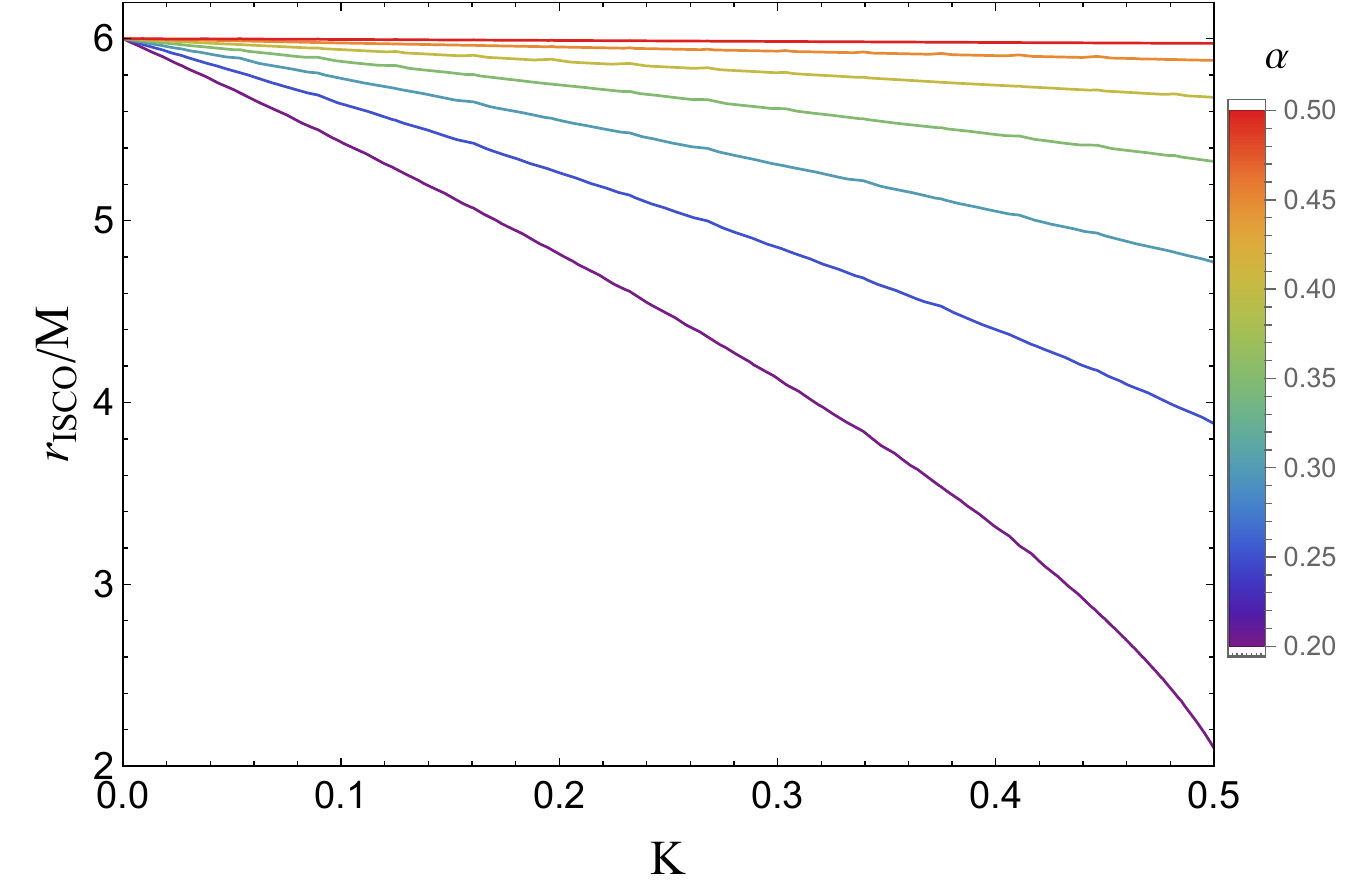}
    \includegraphics[width=0.49\linewidth]{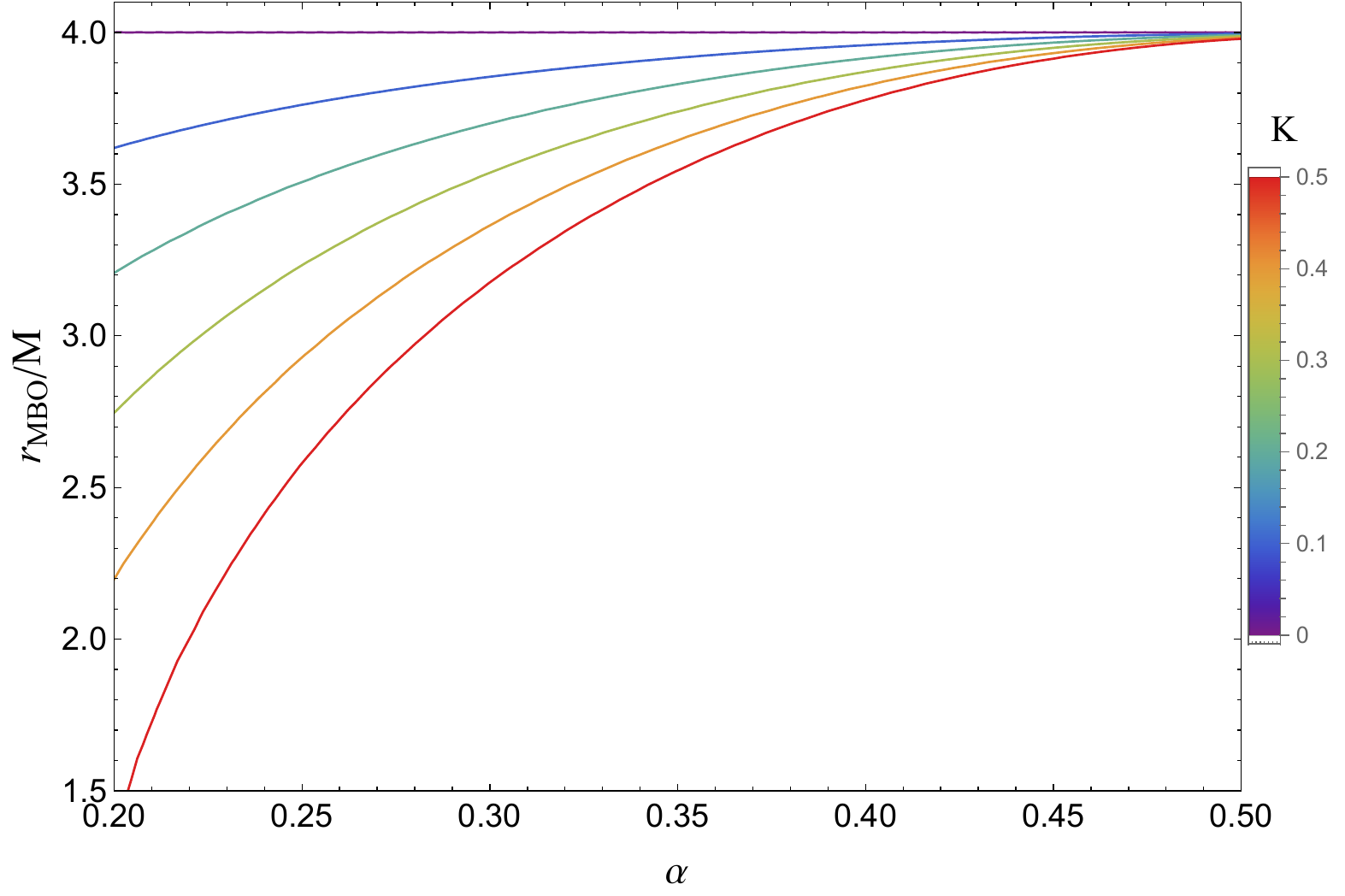}
    \includegraphics[width=0.49\linewidth]{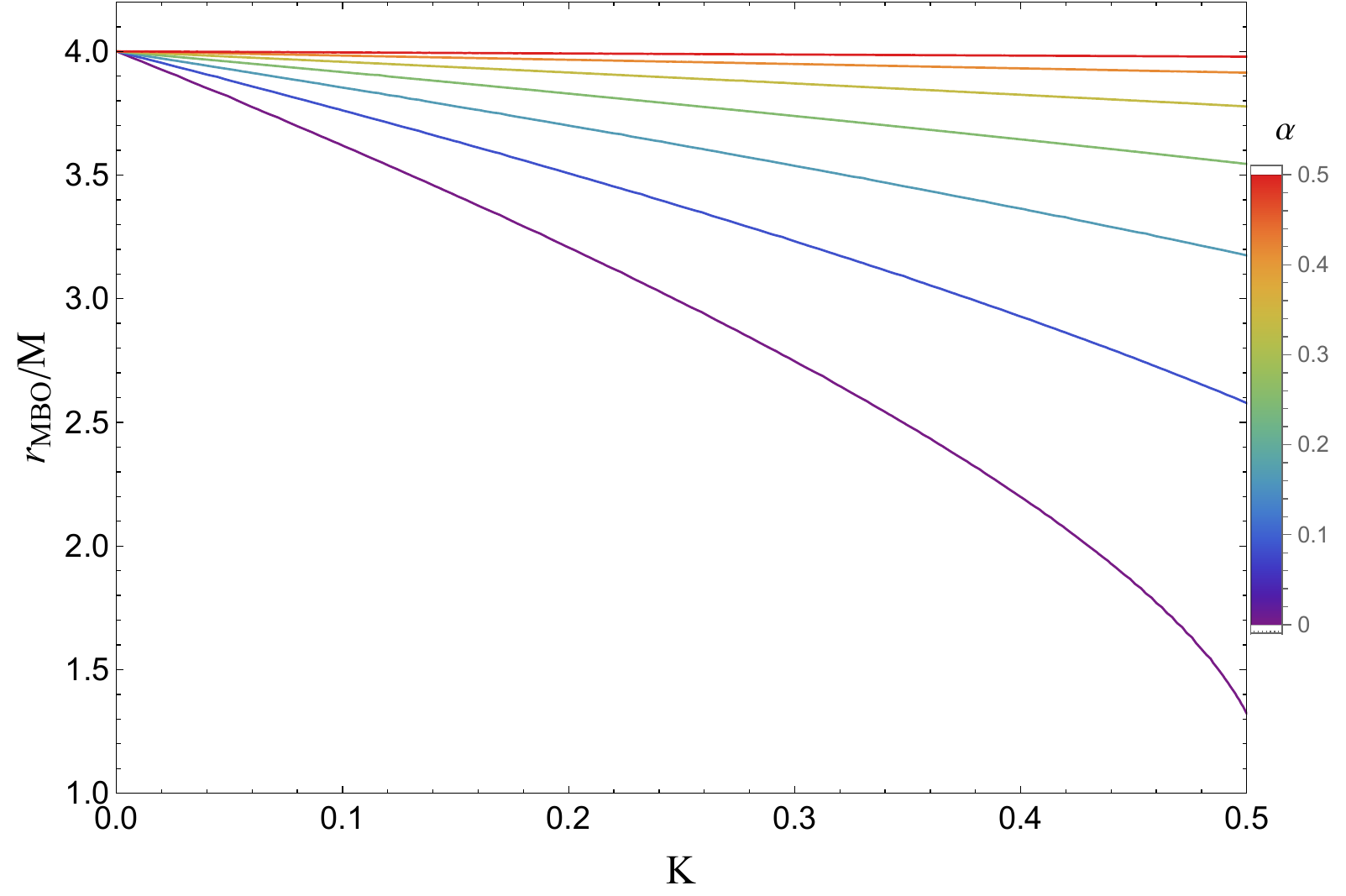}
    \includegraphics[width=0.49\linewidth]{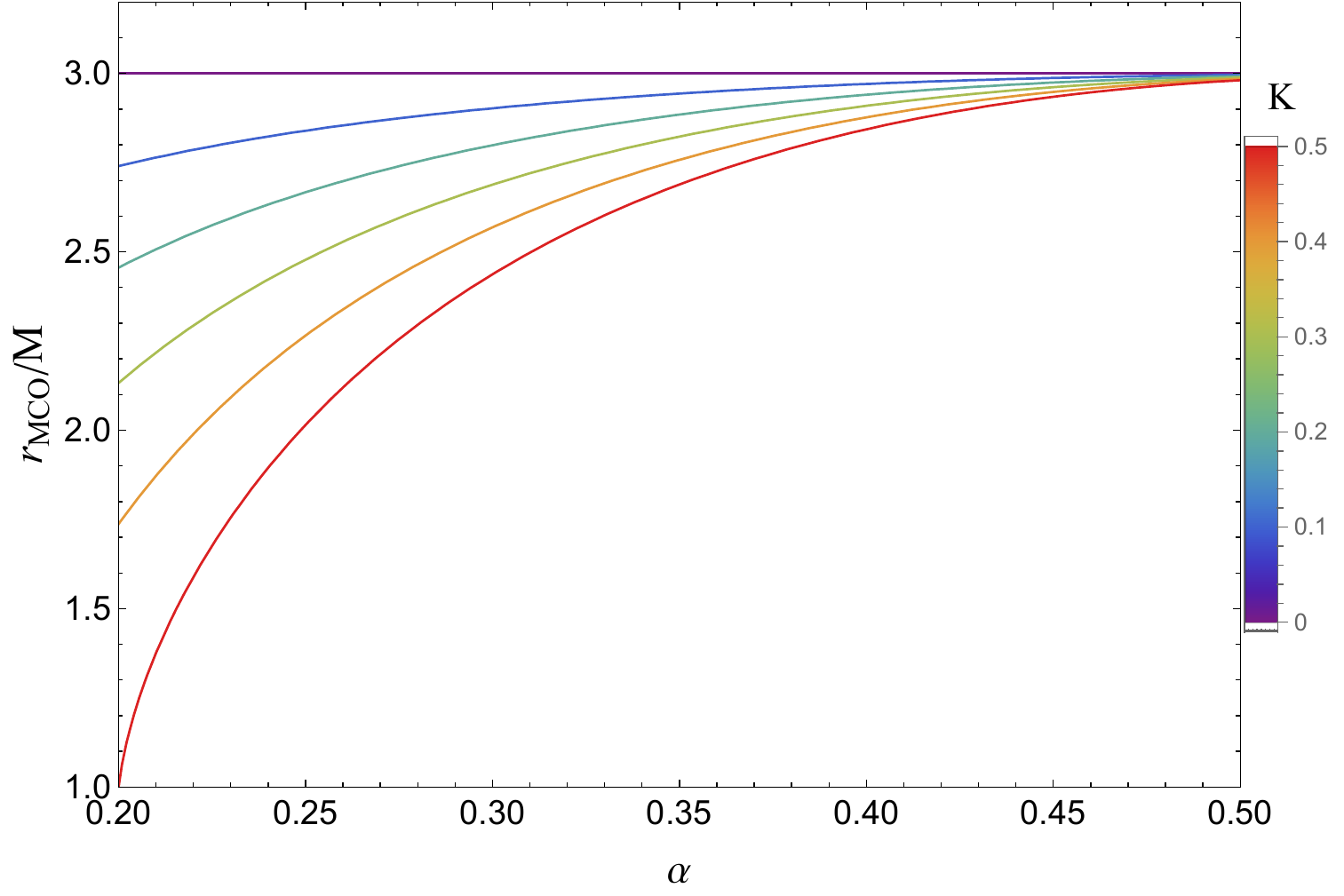}
    \includegraphics[width=0.49\linewidth]{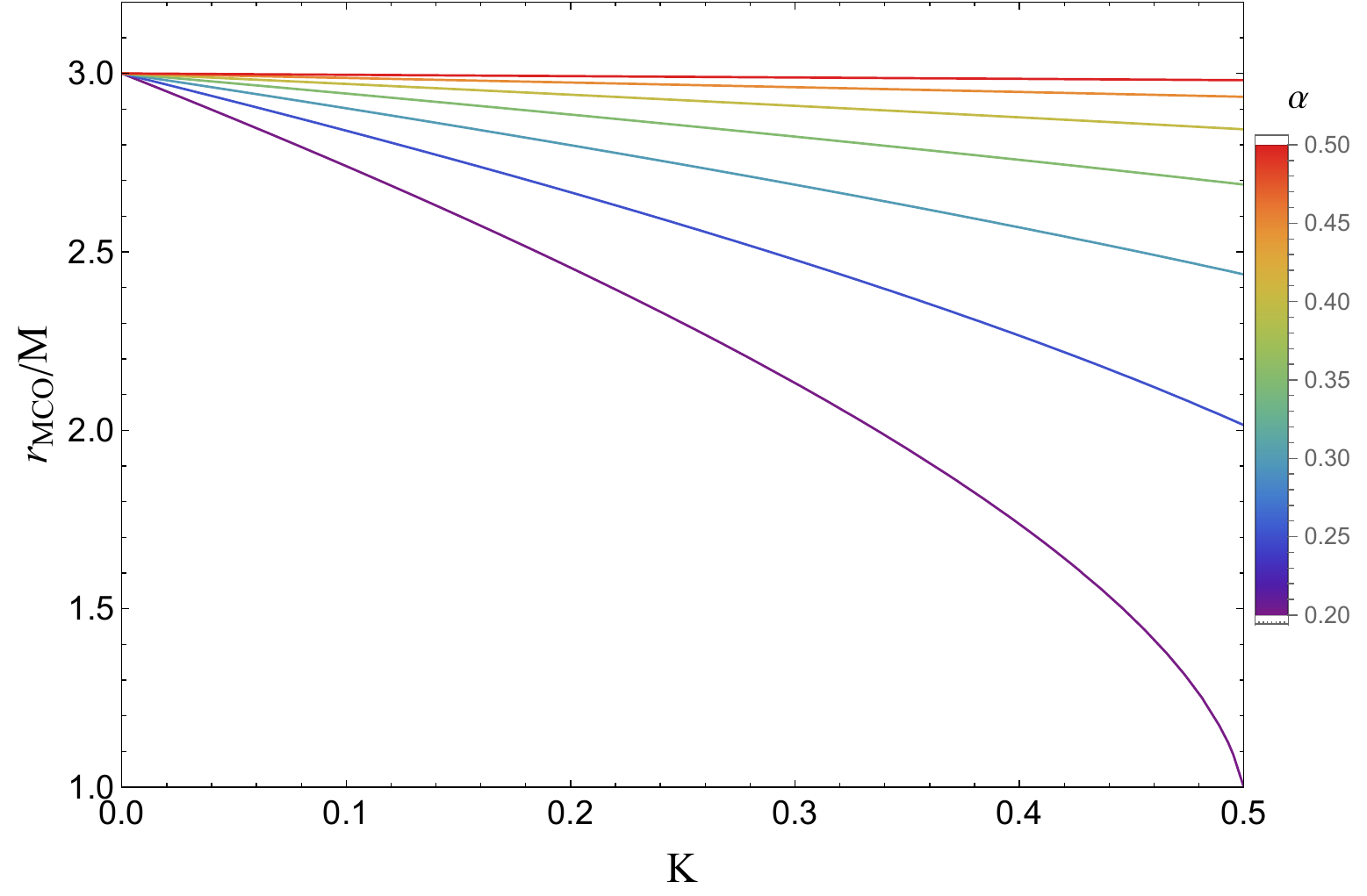}
 \caption{Characteristic circular-orbit radii for the dust-field QFMG black hole. The top, middle, and bottom panels show $r_{\rm ISCO}$, $r_{\rm MBO}$, and $r_{\rm MCO}$, respectively, while the left and right columns display their dependence on $K$ and $\alpha$.}
    \label{f5}
\end{figure*}

Fig.~\ref{f5} (middle panels) shows how $r_{\rm MBO}$, obtained numerically from Eq.~(\ref{eq:mbo_qfmg_simplified}), depends on $\alpha$ and $K$. In the middle-left panel, the Schwarzschild curve $K=0$ stays flat at $r_{\rm MBO}=4M$ for all $\alpha$, confirming that the dust field alone drives the deviation from GR; at fixed $\alpha$, larger $K$ pulls $r_{\rm MBO}$ to smaller radii, while at fixed $K$, larger $\alpha$ shifts it outward, approaching $4M$ as $\alpha\to0.5$. The middle-right panel shows the same trend from the complementary axis: at fixed $\alpha$, increasing $K$ steadily lowers $r_{\rm MBO}$, with a steep decline for small $\alpha$ and an increasingly shallow one as $\alpha\to0.5$, where $r_{\rm MBO}$ stays close to $4M$ even at $K=0.5$. As with the ISCO, $\alpha$ and $K$ shift $r_{\rm MBO}$ in opposite directions -- the QF effect pushing it outward, the dust field pulling it inward.

\subsection{Minimum circular orbit \label{subsec:mco}}

The minimum circular orbit (MCO) marks the innermost radius at which a circular timelike geodesic can exist at all, independent of stability; it is approached only in the limit of divergent specific energy and angular momentum, and in the Schwarzschild case it coincides with the photon sphere. No circular timelike orbit is possible for $r<r_{\rm MCO}$, so it sets the absolute inner boundary of circular orbital motion in this spacetime.

From Eq.~\eqref{eq:Ec_Lc_general}, this radius follows from the vanishing of the common denominator of $\mathcal{E}_c^2$ and $\mathcal{L}_c^2$,
\begin{equation}
2f(r)-rf'(r)=0,
\label{eq:mco_general}
\end{equation}
with $r_{\rm MCO}$ given by its positive root outside the horizon. For the dust-field QFMG metric, this becomes
\begin{equation}
(\alpha-1)K r^{\frac{3\alpha}{3\alpha-2}}
+\alpha(3M-r)=0,
\label{eq:mco_qfmg_explicit}
\end{equation}

which for $K=0$ reduces to the Schwarzschild value $r_{\rm MCO}=3M$.

Figure~\ref{f5} (bottom panels) shows $r_{\rm MCO}$, obtained from Eq.~(\ref{eq:mco_qfmg_explicit}). Being the innermost of the three characteristic radii considered, $r_{\rm MCO}$ probes the region closest to the horizon, where the dust field's influence on the geometry is strongest; consistent with this, its sensitivity to $K$ is the sharpest of the three quantities examined, especially at small $\alpha$. Its qualitative dependence on $K$ and $\alpha$ otherwise follows the same pattern as $r_{\rm ISCO}$ and $r_{\rm MBO}$: the Schwarzschild value $r_{\rm MCO}=3M$ is recovered whenever $K=0$ or $\alpha\to0.5$, and $r_{\rm MCO}$ is drawn inward as the dust field strengthens or the QF effect weakens. Throughout the parameter space explored, the ordering $r_{\rm MCO}<r_{\rm MBO}<r_{\rm ISCO}$ familiar from Schwarzschild spacetime is preserved, indicating that the dust field and the QF effect rescale the strong-field orbital structure without disturbing this ordering.

\section{Fundamental frequencies \label{sec.4}}

In this section, we study the fundamental frequencies associated with the motion of a test particle on a circular geodesic in the dust-field QFMG spacetime.

\subsection{Keplerian frequency}

The Keplerian (azimuthal) angular velocity of a circular geodesic, as measured by a static observer at infinity, is defined as $\Omega_K\equiv d\phi/dt$. For a circular orbit ($\dot r=\ddot r=0$) in the equatorial plane, the radial component of the geodesic equation reduces to 
\begin{eqnarray}
    \Gamma^r_{tt}\dot t^{\,2}+\Gamma^r_{\phi\phi}\dot\phi^{\,2}=0
\end{eqnarray}
which for a static, spherically symmetric metric with $g_{rr}=1/f(r)$ and substituting the metric function Eq.~\eqref{eqBdust} then yields
\begin{equation}
\Omega_K^2=\left(\frac{d\phi}{dt}\right)^2=\frac{\partial_rf(r)}{2r}=\frac{M}{r^3}-\frac{K}{3\alpha}r^{\frac{2}{3\alpha-2}-2}.
\label{eq:OmegaK_qfmg}
\end{equation}

The Schwarzschild result $\Omega_K^2=M/r^3$ is recovered for $K=0$. Fig.~\ref{f6} (left panel) shows the Keplerian frequency $\Omega_K$ as a function of $r/M$, obtained from Eq.~(\ref{eq:OmegaK_qfmg}). Since the QF-effect term enters $\Omega_K^2$ with a negative sign, the dust field systematically pulls $\Omega_K$ below its Schwarzschild value at the same radius: at fixed $\alpha$, a larger $K$ lowers $\Omega_K$ further, while at fixed $K$, a larger $\alpha$ weakens this suppression and brings the curve closer to Schwarzschild. As already found for $r_{\rm MCO}$ (Fig.~5, bottom panels), a stronger dust field also shifts the minimum circular-orbit radius inward, so the QFMG curves extend to smaller $r$ than the Schwarzschild curve, which terminates at $r_{\rm MCO}=3M$. At large $r$, all curves converge onto the Schwarzschild case, as the influence of $K$ and $\alpha$ fades far from the black hole.

\subsection{Epicyclic frequencies}

A test particle on a stable circular orbit in the equatorial plane of the dust-field QFMG black hole oscillates along the radial and vertical directions when slightly displaced from that orbit, $r_0+\delta r$ and $\pi/2+\delta\theta$. These small oscillations obey the harmonic-oscillator equations
\begin{eqnarray}
\frac{d^2\delta r}{dt^2}+\Omega_r^2\,\delta r=0, \qquad \frac{d^2\delta\theta}{dt^2}+\Omega_\theta^2\,\delta\theta=0,
\end{eqnarray}
where
\begin{eqnarray}
&&\Omega_r^2=-\frac{1}{2g_{rr}(u^t)^2}\partial_r^2V_{\rm eff}(r,\theta)\Big|_{\theta=\pi/2},\\
&&\Omega_\theta^2=-\frac{1}{2g_{\theta\theta}(u^t)^2}\partial_\theta^2V_{\rm eff}(r,\theta)\Big|_{\theta=\pi/2},
\end{eqnarray}
are the radial and vertical angular frequencies, respectively. For a static, spherically symmetric spacetime these reduce to \citep{Rayimbaev2021Galax}
\begin{eqnarray}
\Omega_\theta^2=\Omega_K^2, \qquad
\Omega_r^2=\Omega_K^2\left[\left(3+\frac{r\partial_{rr}f(r)}{\partial_rf(r)}\right)f(r)-2r\partial_rf(r)\right].
\label{eq:Omegar_general}
\end{eqnarray}
Substituting the dust-field QFMG metric function then gives
\begin{align}
\Omega_r^2
=\frac{M}{r^3}
-\frac{6M^2}{r^4}
-\frac{2K(3\alpha-1)}{3\alpha(3\alpha-2)}
r^{\frac{2}{3\alpha-2}-2}
-\frac{KM(3\alpha^2-16\alpha+8)}{\alpha(3\alpha-2)}
r^{\frac{2}{3\alpha-2}-3}
+\frac{2K^2(\alpha-1)}{3\alpha^2}
r^{\frac{4}{3\alpha-2}-2}.
\label{eq:omega_r_qfmg}
\end{align}

\begin{figure*}
\centering \includegraphics[width=0.49\linewidth]{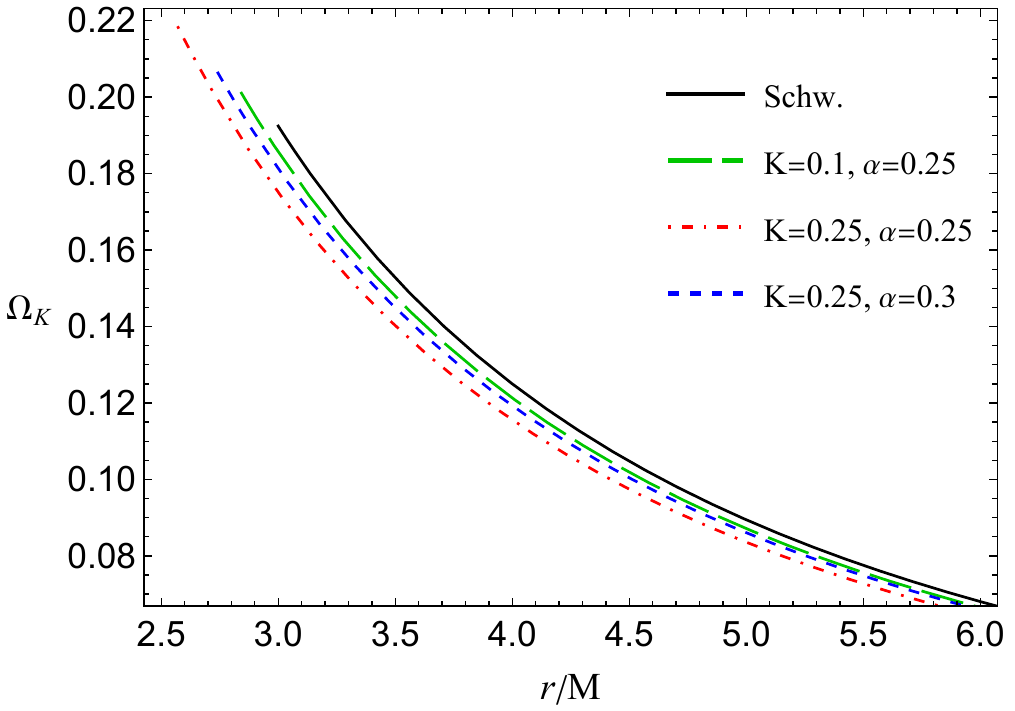}
\includegraphics[width=0.49\linewidth]{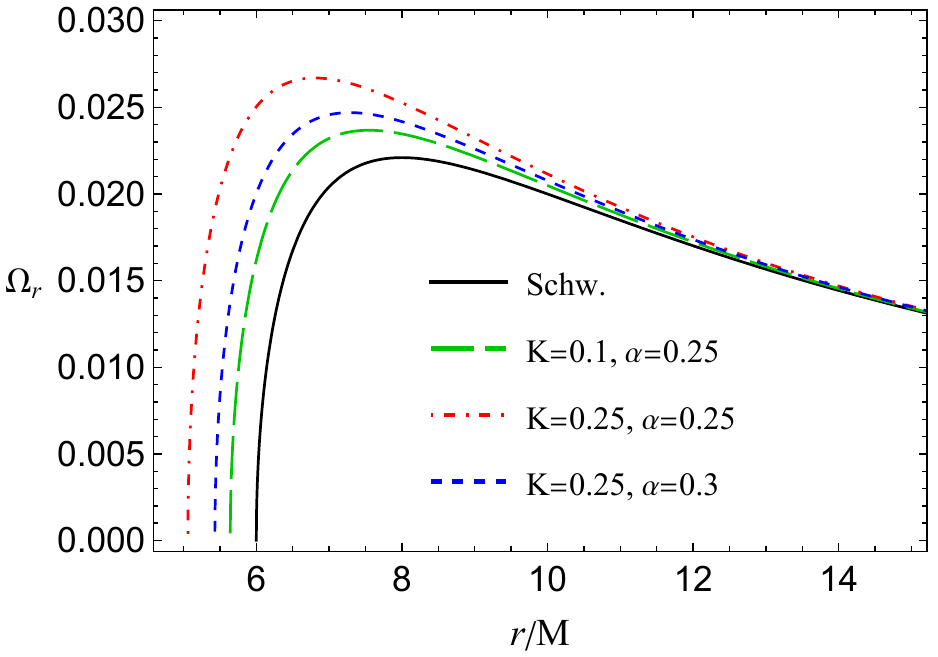}
\caption{Keplerian frequency $\Omega_K$ (left panel) and radial epicyclic frequency $\Omega_r$ (right panel) for the dust-field QFMG black hole}
\label{f6} 
\end{figure*}

Figure~\ref{f6} (right panel) shows the radial epicyclic frequency $\Omega_r$ as a function of radial coordinate. Each curve starts at $\Omega_r=0$, since by definition this radius is precisely $r_{\rm ISCO}$: for $r<r_{\rm ISCO}$ a circular orbit is unstable and no restoring force exists. Moving outward from the ISCO, $\Omega_r$ rises to a maximum, reflecting a radial range where the radial restoring force is strongest; beyond this peak it decreases monotonically, as the weakening gravitational pull at large $r$ causes all orbital frequencies to fall off toward their common weak-field, nearly Newtonian behavior.

At fixed $\alpha$, increasing $K$ pulls the zero of $\Omega_r$ inward -- consistent with the inward shift of $r_{\rm ISCO}$ and simultaneously raises the peak and shifts it to smaller $r$, indicating that a stronger dust field tightens the radial restoring force and allows stable oscillatory motion closer to the black hole. At fixed $K$, increasing $\alpha$ has the opposite effect: it pushes the zero of $\Omega_r$ outward, lowers the peak, and shifts it to larger $r$, consistent with the weakened near-horizon confinement found for $V_{\rm eff}$ in Fig.~\ref{V.eff} and the outward displacement of $r_{\rm ISCO}$ in Fig.~\ref{f5}. At large $r$, all curves converge onto the Schwarzschild profile, as the dust field and the QF effect become dynamically irrelevant far from the black hole. 

\section{Quasiperiodic oscillations}
\label{sec:qpo}

Quasi-periodic oscillations (QPOs) were first detected in the X-ray flux of an accreting compact object in 1985, when van der Klis et al. reported a broad, intensity-dependent variability feature in the 20--40~Hz range from the neutron-star binary GX~5$-$1 \citep{1985Natur.316..225V}. A decade later, the launch of the Rossi X-ray Timing Explorer (RXTE) opened access to much higher frequencies, leading to the discovery of kilohertz (kHz) QPOs, first in Scorpius X-1 \citep{1996ApJ...469L...1V} and shortly afterward in several other neutron-star low-mass X-ray binaries \citep{1996ApJ...469L...9S}. These early studies established QPOs as a common and remarkably well-defined feature of accreting compact objects, and subsequent observational and theoretical work has substantially expanded the sample and the frequency range over which they are detected \citep{2006csxs.book...39V,2006ARA&A..44...49R}.

Depending on the source state and the instrument sensitivity, QPOs appear either as a single peak in the power spectrum or as two simultaneous peaks, commonly labeled the upper ($\nu_U$) and lower ($\nu_L$) frequencies. The latter, so-called twin-peak or double-peak QPOs, are of particular diagnostic value, since the ratio $\nu_U/\nu_L$ and the individual frequencies both carry information about the underlying spacetime and are frequently found to cluster near small-integer ratios such as $3{:}2$ \citep{2006ARA&A..44...49R}. Ratios of $4{:}3$ and $5{:}4$ have also been reported, for example in the atoll neutron-star sources \citep{2010MNRAS.401.1290B} and in the ultraluminous X-ray source NGC~5408~X$-$1 \citep{2007ApJ...660..580S}.

In stellar-mass BH X-ray binaries, twin-peak high-frequency QPOs were first firmly established in the Galactic microquasar GRO~J1655$-$40, where Strohmayer detected a $450$~Hz QPO occurring simultaneously with a previously known $\sim300$~Hz QPO, providing the first evidence of a pair of high-frequency QPOs from a BH. A twin-peak pair was subsequently reported in the microquasar GRS~1915$+$105 \citep{2001ApJ...554L.169S}, and both sources remain among the best-studied stellar-mass BH twin-peak QPO systems to date \citep{2006ARA&A..44...49R}.

Twin-peak QPOs have also been identified in ultraluminous X-ray sources hosting intermediate-mass BH candidates, where the frequencies scale down in proportion to the larger BH mass. Pasham, Strohmayer \& Mushotzky discovered a stable $3{:}2$ twin-peak QPO pair at $3.32$ and $5.07$~Hz in M82~X$-$1, from which they inferred a BH mass of several hundred solar masses \citep{2014Natur.513...74P}. A comparable twin-peak pair was subsequently found in NGC~1313~X$-$1 \citep{2015ApJ...811L..11P}, extending the twin-peak QPO phenomenology to the intermediate-mass regime and providing the four sources used in our own MCMC analysis in next section.

Several theoretical models have been proposed to explain the observed twin-peak frequencies and their preferred ratios. The RP model identifies the upper frequency with the orbital frequency and the lower frequency with the periastron precession frequency of a slightly eccentric orbit \citep{1998ApJ...492L..59S,1999PhRvL..82...17S}. The warped-disk (WD) model instead relates the twin peaks to non-axisymmetric oscillation modes of a warped thin accretion disk \citep{2004PASJ...56..905K,2008PASJ...60..111K}, while the epicyclic-resonance (ER) model marks them to a parametric or forced resonance between the radial and vertical epicyclic frequencies at a specific radius \citep{2001A&A...374L..19A}. A further class of models, diskoseismology, associates the QPOs with global oscillation modes trapped within the accretion disk itself \citep{1999PhR...311..259W}. These frameworks differ in which orbital or disk frequencies they identify with the observed peaks, and consequently predict different mass and spacetime-parameter dependences for a given twin-peak pair.

The RP, WD and ER models have each been applied extensively to test modified and charged black hole solutions against observed twin-peak QPOs, including charged black holes in Einstein--Maxwell-scalar theory \citep{2022EPJC...82.1110R}, Kiselev black holes with a cloud of strings \citep{2024ChPhC..48e5104R}, magnetized black holes in braneworlds \citep{2024EPJC...84.1114R}, charged particles in Simpson--Visser spacetime \citep{2023EPJC...83..854V}, magnetized black holes in Bertotti--Robinson geometry \citep{2025EPJC...85.1017S}, magnetized noncommutative-inspired black holes \citep{2026PhLB..88040821S}, Einstein-nonlinear-Maxwell-Yukawa black holes through their gravitational-wave and QPO signatures \citep{2025EPJC...85.1340Z,2026NuPhB102617432S}, hairy black holes in Horndeski gravity \citep{2023EPJC...83..572R}.

In this work, we compare all three of these models -- RP, WD, and ER4 -- using the twin-peak QPO frequencies from the Keplerian and radial epicyclic frequencies derived in Sec.~\ref{sec.4}. The angular frequencies are converted to Hz as follow:
\begin{equation}
\nu_i=\frac{1}{2\pi}\frac{c^3}{GM}\,\Omega_i,
\qquad i=K,r ,
\label{eq:frequency_conversion}
\end{equation}
where $M$ is the black hole mass. In the RP model, the upper and lower frequencies are
\begin{equation}
\nu_U=\nu_K,\qquad
\nu_L=\nu_K-\nu_r .
\label{eq:rp_model}
\end{equation}
The WD model is defined as
\begin{equation}
\nu_U=2\nu_K-\nu_r,\qquad
\nu_L=2(\nu_K-\nu_r).
\label{eq:wd_model}
\end{equation}
For the ER4 models used here, we take
\begin{equation}
\nu_U=\nu_K+\nu_r,\qquad
\nu_L=\nu_K-\nu_r ,
\label{eq:er4_model}
\end{equation}

\begin{figure*}
\centering
\includegraphics[width=0.46\linewidth]{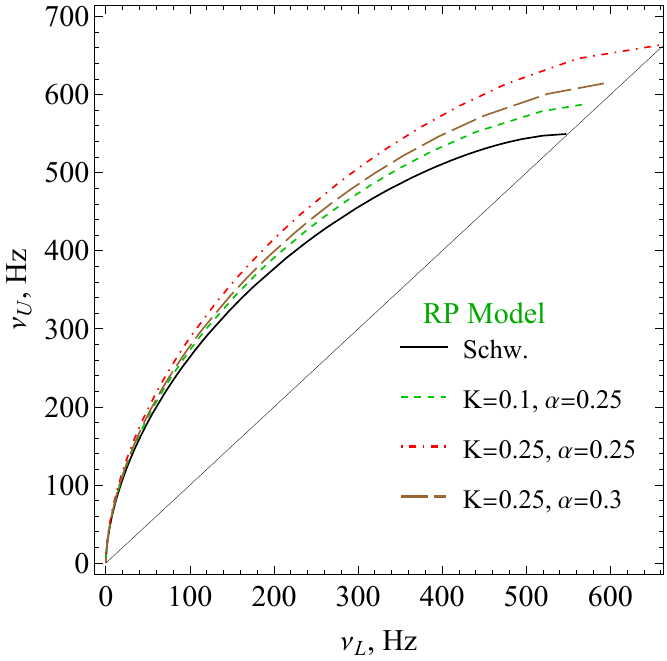}
\includegraphics[width=0.48\linewidth]{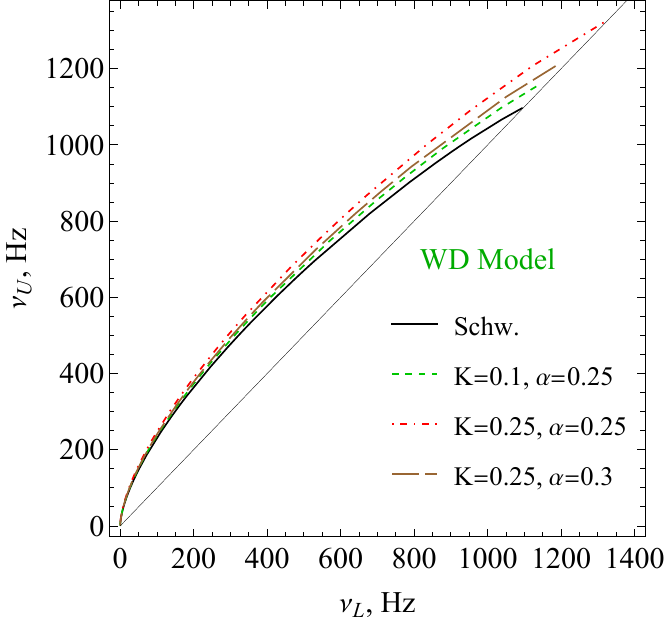}
\includegraphics[width=0.46\linewidth]{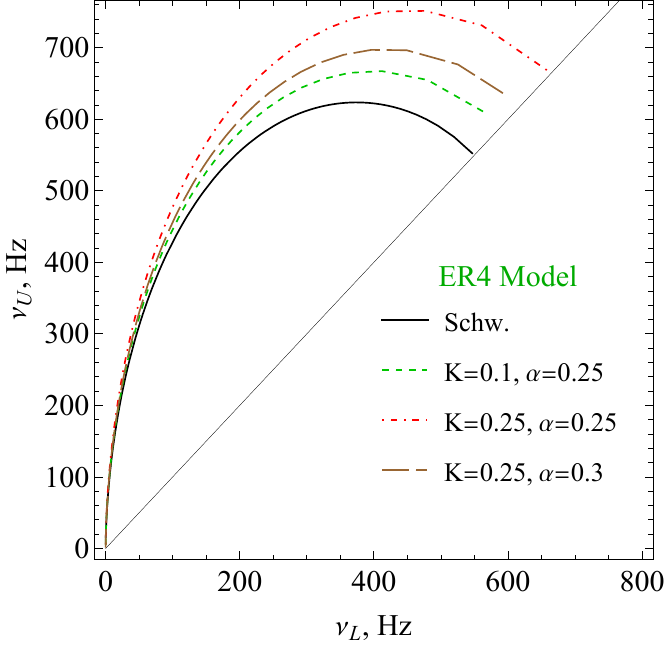}
\caption{Upper and lower QPO frequencies for the dust-field QFMG black hole in the RP, WD and ER4 models.}
\label{f7}
\end{figure*}

Fig.~\ref{f7} shows the $\nu_L$--$\nu_U$ relations for the dust-field QFMG black hole in the RP, WD, and ER4 models.

\begin{itemize}

\item \textit{Low-frequency regime (large orbital radii):} Here the QPO orbits lie far from the black hole, where the gravitational field is weak and both $\nu_K$ and $\nu_r$ are small. In all three models, the modified curves stay close to the Schwarzschild line, showing that the dust field and the QF effect produce a negligible frequency shift at large radii -- consistent with the asymptotic flatness of the spacetime, where the metric approaches the Schwarzschild form and the influence of $K$ and $\alpha$ drops.

\item \textit{Intermediate regime:} As the QPO orbit moves inward, the dust field and the QF effect become dynamically relevant. In the RP and WD models, the modified curves separate from the Schwarzschild track and shift upward, so that for a given $\nu_L$ the corresponding $\nu_U$ is larger than in the Schwarzschild case; a larger $K$ at fixed $\alpha$ strengthens this upward shift, while a larger $\alpha$ at fixed $K$ partially offsets it. In the ER4 model, the curve instead develops a turnover, whose peak height and location shift relative to the Schwarzschild case as $K$ and $\alpha$ are varied.

\item \textit{Near-ISCO regime (highest frequencies attainable):} Here the separation from the Schwarzschild curve is most pronounced. In the RP and WD models, all modified curves remain above the Schwarzschild line, with the upward displacement growing with $K$ and being partially reduced by a larger $\alpha$; since these frequencies correspond to orbits close to the inner edge of the stable circular region, this regime offers the most sensitive window for constraining $K$ and $\alpha$ from twin-peak QPO data. In the ER4 model, each curve terminates at the ISCO, and the maximum frequency it reaches shifts with $K$ and $\alpha$ in a manner consistent with the displacement of $r_{\rm ISCO}$ found in the top panels of Fig.~\ref{f5}.

\end{itemize}

\subsection{Radius of QPO orbit}

In this subsection, we determine the radii at which the RP, WD, and ER4 frequencies fall into the small-integer ratios discussed above, and compare them to the ISCO radius. A resonant radius is a physically motivated benchmark because, for a given ratio $m{:}n$, it depends only on the spacetime and not on the observed frequencies themselves, and consequently scales simply with the black hole mass \citep{2001A&A...374L..19A}. Comparing it to the ISCO is essential for two reasons: first, since $\nu_r\to0$ exactly at the ISCO, the upper and lower frequencies necessarily merge into a single peak there, so a genuine twin-peak signal requires the resonant orbit to lie strictly outside the ISCO; second, the location of the resonant radius relative to the ISCO has itself been used to probe the structure of the inner accretion disc in Keplerian discs around compact objects \citep{2011A&A...525A..82S}. We compute the resonant radii for the $3{:}2$, $4{:}3$, and $5{:}4$ ratios in each of the three models and display them together with the ISCO radius as functions of $\alpha$ and $K$.

\begin{figure*}
\centering  
\includegraphics[width=0.32\linewidth]{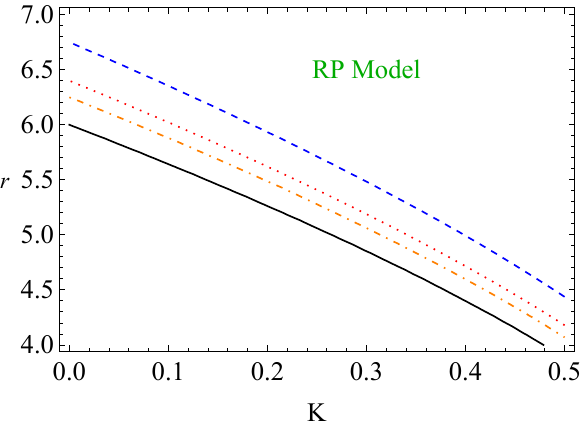}
\includegraphics[width=0.31\linewidth]{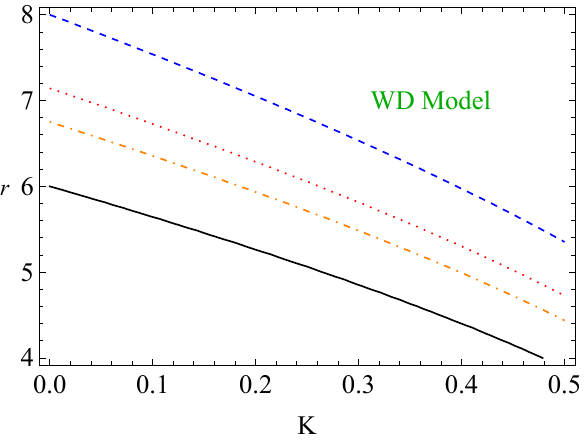}
\includegraphics[width=0.32\linewidth]{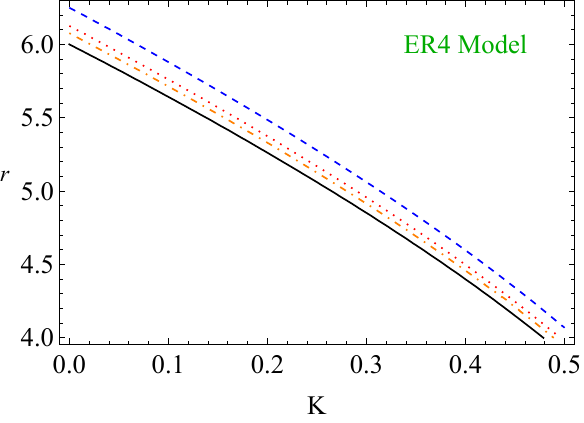}
\caption{Resonance radii for the RP, WD, and ER4 models as functions of the dust-field strength $K$. Here, black solid, blue dashed, red dotted, and orange dot--dashed lines correspond to ISCO, $\nu_U/\nu_L=3:2$, $\nu_U/\nu_L=4:3$, and $\nu_U/\nu_L=5:4$ resonance radii, and we select a black hole mass of $M=6M_\odot$ at fixed $\alpha=0.25$.}
\label{f8}
\end{figure*}

\begin{figure*}[htbp]
\centering  
\includegraphics[width=0.32\linewidth]{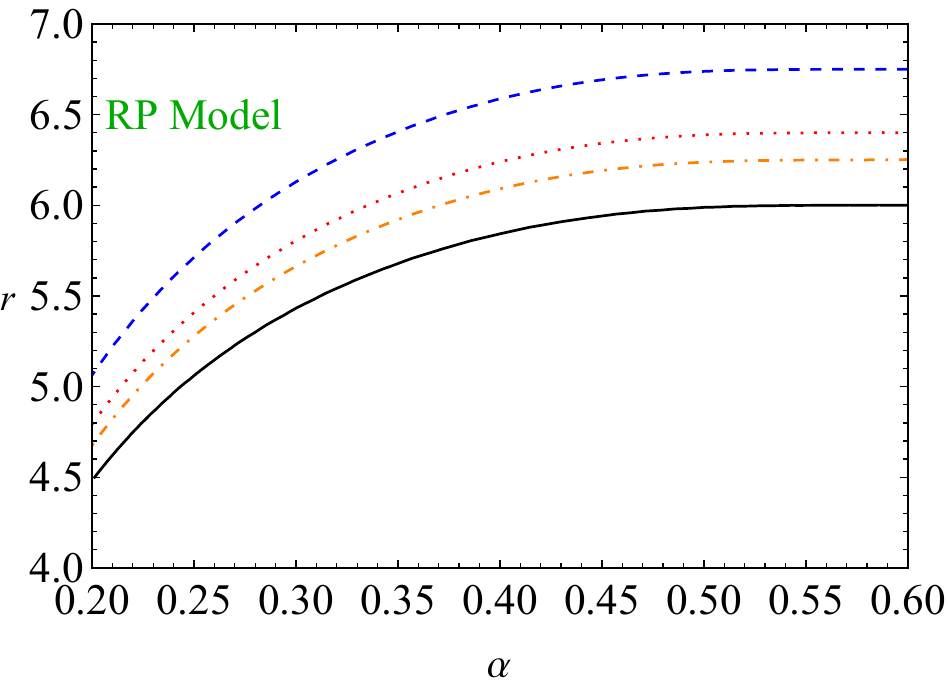}
\includegraphics[width=0.32\linewidth]{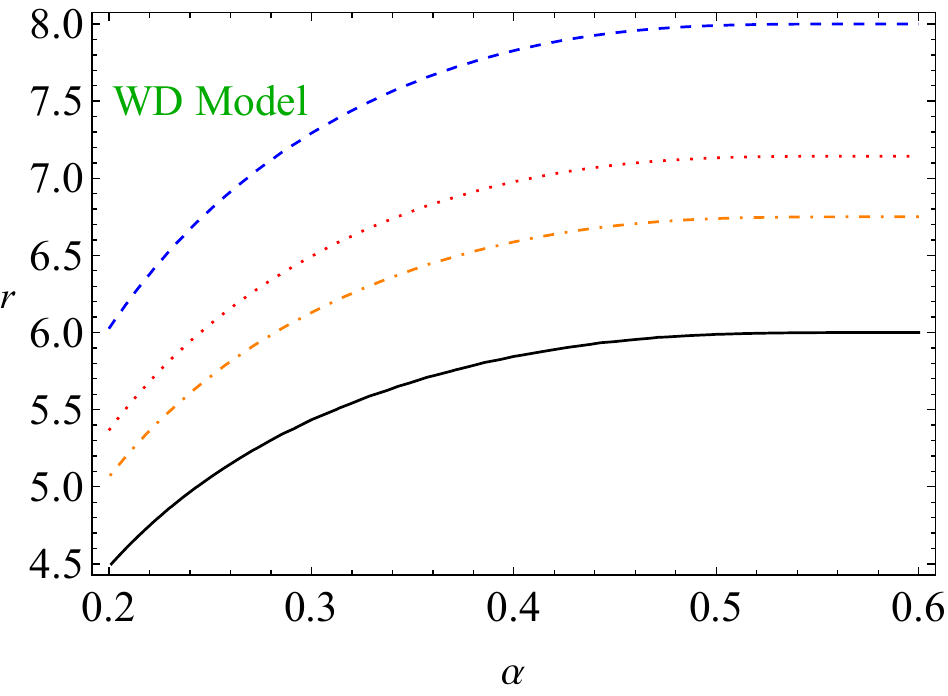}
\includegraphics[width=0.32\linewidth]{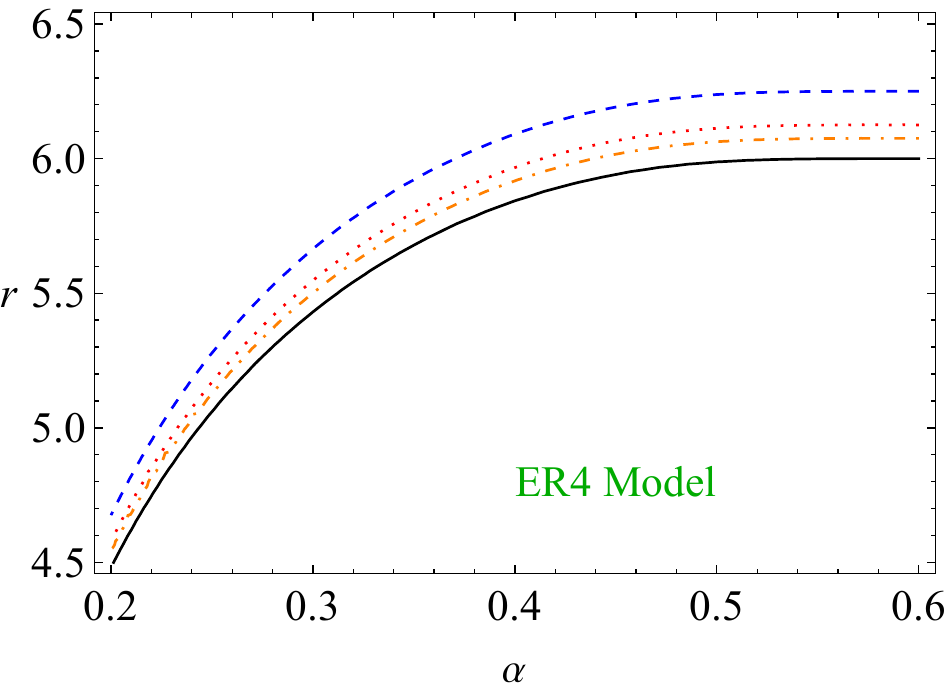}

\caption{Similar as Fig. \ref{f8}, but $r$ depends on $\alpha$. Here $K=0.25$.}
\label{f9}
\end{figure*}

Figure~\ref{f8} shows the resonance radii for the $3{:}2$, $4{:}3$, and $5{:}4$ ratios together with $r_{\rm ISCO}$ as functions of $K$, in the RP (left), WD (middle), and ER4 (right) models. In all three models, increasing $K$ lowers every resonance radius and the ISCO, so a stronger dust field draws the entire resonant structure inward together, with each model's set of curves closing in on $r=4$--$4.5$ near $K\simeq0.5$. The separation between the resonance curves $\Delta r$ is model-dependent: it is largest in the WD model, intermediate in the RP model, and smallest in the ER4 model, so the choice of QPO model itself sets how sharply the $3{:}2$, $4{:}3$, and $5{:}4$ radii are spread out at a given $K$. Within each model, this separation also follows the frequency ratio itself: as $\nu_U/\nu_L\to1$, the resonance radius approaches $r_{\rm ISCO}$, so the $5{:}4$ curve lies closest to the ISCO line, the $4{:}3$ curve slightly farther, and the $3{:}2$ curve farthest away, in all three panels.

Figure~\ref{f9} shows the resonance radii as functions of $\alpha$. In all three models, increasing $\alpha$ raises every resonance radius together with $r_{\rm ISCO}$, consistent with the anti--gravitating character of the QF effect: strengthening it pushes the resonant structure outward to larger radii. This outward shift is steep over $\alpha\in(0.2,0.4)$, where  increase in $\alpha$ already produces a noticeable change in $r$, but flattens by changing over $\alpha\in(0.4,0.6)$, where all curves close to their Schwarzschild values and the QF effect becomes negligible. The relative spacing of the curves -- $\Delta r_{\text{WD}}>\Delta r_{\text{RP}}>\Delta r_{\text{ER4}}$ -- and the convergence of each curve toward $r_{\rm ISCO}$ as $\nu_U/\nu_L\to1$ found for the $K$-dependence in Fig.~\ref{f8}. Since this ordering keeps under variation of spacetime parameters, it appears to be set by the algebraic form of each QPO model rather than by the dust-field QFMG spacetime.

\section{MCMC constraints from twin-peak QPOs \label{section6}}

In this section, we constrain the parameters of the dust-field QFMG black hole using observed twin-peak QPO frequencies, applying the RP, WD, and ER4 models introduced in Sec.~\ref{sec:qpo} separately.

\subsection{Observational data and model parameters}

The observational data used in the MCMC analysis are listed in Tab.~\ref{table1}. The sample includes the Galactic microquasars XTE J1550$-$564 and GRO J1655$-$40, together with the ultraluminous X-ray sources M82 X$-$1 and NGC 1313 X$-$1, both regarded as candidate intermediate-mass black holes. For each source, we use the observed upper and lower QPO frequencies, $\nu_U^{\rm obs}$ and $\nu_L^{\rm obs}$, together with their corresponding uncertainties. The mass values listed in Tab.~\ref{table1} are given only for comparison with independent observational estimates; they are not imposed as priors in the present sampling, so the constraints are obtained directly from the twin-peak QPO frequencies themselves.

The model is described by the parameter vector $\vartheta=\left(M,K,\alpha,r\right)$ where $M$ is the black hole mass, $K$ is the dust-field strength, $\alpha$ is the QF parameter, and $r$ is the orbital radius at which the observed QPO pair is generated.

For each sampled parameter set, the theoretical upper and lower frequencies, $\nu_U^{\rm th}(\vartheta)$ and $\nu_L^{\rm th}(\vartheta)$, are evaluated from the Keplerian and radial epicyclic frequencies derived in above section. The MCMC procedure therefore constrains the black hole mass, the dust-field and QF parameters, and the QPO emission radius from the observed twin-peak QPO data, independently for each model.

\begin{table*}[ht!]
\centering
\caption{Frequencies of twin-peak QPOs in microquasars}
\label{table1}
\renewcommand{\arraystretch}{1.2}
\begin{tabular}{| l || c c | c c |}
\hline
Source 
& $\nu_{\rm U}$ [Hz] 
& $\Delta\nu_{\rm U}$ [Hz] 
& $\nu_{\rm L}$ [Hz] 
& $\Delta\nu_{\rm L}$ [Hz] \\
\hline
\hline
XTE~J1550$-$564 \cite{2002ApJ...580.1030R} 
& 276 & $\pm\,3$ & 184 & $\pm\,5$ \\
GRO~J1655$-$40 \cite{2001ApJ...552L..49S}    
& 451 & $\pm\,5$ & 298 & $\pm\,4$ \\
M82 X$-$1 \cite{2014Natur.513...74P}  & 5.07 & $\pm\,0.06$ & 3.32& $\pm\,0.06$  \\
NGC 1313 X$-$1 \cite{2015ApJ...811L..11P}  & 0.46 & $\pm\,0.02$ & 0.29& $\pm\,0.01$\\
\hline
\end{tabular}
\end{table*}

\subsection{Bayesian setup and likelihood function}

We estimate parameters within a Bayesian framework. This approach is useful because the theoretical QPO frequencies depend nonlinearly on the black hole mass, the dust-field strength, the QF parameter, and the QPO orbital radius, and different combinations of these parameters may reproduce similar observed upper and lower QPO frequencies; we therefore explore the full multidimensional parameter space using MCMC simulations rather than a simple best-fit procedure.

For a given source and a given QPO model, the posterior probability distribution is written as
\begin{equation}
{\cal P}(\vartheta|{\cal D},{\cal M})
=
\frac{
{\cal L}({\cal D}|\vartheta,{\cal M})\,
\pi(\vartheta|{\cal M})
}{
{\cal Z}({\cal D}|{\cal M})
},
\label{eq:posterior_qfmg}
\end{equation}
where ${\cal D}$ denotes the observed twin-peak QPO data, ${\cal M}$ denotes the QPO under consideration, ${\cal L}({\cal D}|\vartheta,{\cal M})$ is the likelihood function, $\pi(\vartheta|{\cal M})$ is the prior distribution, and ${\cal Z}({\cal D}|{\cal M})$ is the Bayesian evidence, which acts only as a normalization factor here.

Assuming independent Gaussian observational errors for the upper and lower QPO frequencies, the log-likelihood is written as
\begin{equation}
\ln {\cal L}(\vartheta)
=
\ln {\cal L}_{\rm U}
+
\ln {\cal L}_{\rm L},
\label{eq:loglike_total_qfmg}
\end{equation}
where
\begin{equation}
\ln {\cal L}_{\rm U}
=
-\frac{1}{2}
\frac{
\left[
\nu_{\rm U}^{\rm obs}
-
\nu_{\rm U}^{\rm th}(\vartheta)
\right]^2
}{
\sigma_{\rm U}^{2}
},\quad
\ln {\cal L}_{\rm L}
=
-\frac{1}{2}
\frac{
\left[
\nu_{\rm L}^{\rm obs}
-
\nu_{\rm L}^{\rm th}(\vartheta)
\right]^2
}{
\sigma_{\rm L}^{2}
},
\label{eq:loglike_upper_qfmg}
\end{equation}
with $\nu_{\rm U}^{\rm th}(\vartheta)$ and $\nu_{\rm L}^{\rm th}(\vartheta)$ evaluated from whichever of the RP, WD, or ER4 models of Sec.~\ref{sec:qpo} is being tested. Equivalently, the likelihood can be expressed through the chi-square function,
\begin{equation}
\chi^2(\vartheta)
=
\frac{
\left[
\nu_{\rm U}^{\rm obs}
-
\nu_{\rm U}^{\rm th}(\vartheta)
\right]^2
}{
\sigma_{\rm U}^{2}
}
+
\frac{
\left[
\nu_{\rm L}^{\rm obs}
-
\nu_{\rm L}^{\rm th}(\vartheta)
\right]^2
}{
\sigma_{\rm L}^{2}
}.
\label{eq:chi_square_qfmg}
\end{equation}
Thus, the likelihood measures how well a given parameter set reproduces the observed twin-peak QPO frequencies, separately for each of the RP, WD, and ER4 models.

The posterior distribution is sampled using the affine-invariant ensemble sampler implemented in the Python package \texttt{emcee} \citep{2013PASP..125..306F}. For each source and model, the sampler generates a chain of trial parameter sets and evaluates the posterior probability at each step. After discarding the burn-in part and applying thinning, the remaining samples are used to obtain the marginalized posterior distributions. The quoted parameter constraints correspond to the median values and the associated $1\sigma$ credible intervals, allowing us to estimate not only the best-fit values but also the uncertainties and correlations among $M$, $K$, $\alpha$, and $r$.

\subsection{Prior distributions}

In this subsection, we describe how we obtain the prior distributions used in the MCMC analysis. Since the theoretical QPO frequencies depend nonlinearly on the black hole mass, the dust-field strength, the QF parameter, and the QPO radius, the prior ranges cannot be chosen arbitrarily. They must satisfy the basic physical requirements of the model: the spacetime must admit a black hole horizon, the circular orbit must be stable, and the theoretical frequencies of the model under consideration must be able to reproduce the observed twin-peak QPO data.

Only the random-scan points satisfying
\begin{equation}
\chi^2(\vartheta)<1,
\qquad
r_{\rm QPO}\geq r_{\rm ISCO}
\label{eq:prior_selection_conditions}
\end{equation}
are accepted for obtaining the priors. The first condition selects parameter sets that reproduce the observed QPO frequencies with good accuracy, while the second condition guarantees that the QPO orbit is located in the stable circular region outside the ISCO.

We choose the initial scan intervals to cover the physically relevant parameter space: the ranges of $K$ and $\alpha$ are guided by the horizon-existence analysis of Fig.~\ref{fig:bh-nbh}, while the QPO radius is restricted to the region where stable circular motion is possible, with $r_{\rm QPO}\geq r_{\rm ISCO}$ imposed on every accepted point. We find more than $100$ accepted values for each source and model, and use them to determine the Gaussian prior parameters. For each parameter $\vartheta\in\{M,K,\alpha,r\}$, given $N_{\rm acc}$ accepted values $\vartheta_1,\vartheta_2,\ldots,\vartheta_{N_{\rm acc}}$, the central value of the Gaussian prior is the arithmetic mean,
\begin{equation}
\mu(\vartheta)=
\frac{1}{N_{\rm acc}}
\sum_{j=1}^{N_{\rm acc}} \vartheta_j ,
\label{eq:prior_mu_definition}
\end{equation}
and its width is the mean absolute deviation from this value,
\begin{equation}
\sigma(\vartheta)=
\frac{1}{N_{\rm acc}}
\sum_{j=1}^{N_{\rm acc}}
\left|\mu(\vartheta)-\vartheta_j\right| ,
\label{eq:prior_sigma_definition}
\end{equation}
so that the Gaussian prior for each parameter is written as
\begin{equation}
\pi(\vartheta)
\propto
\exp\left[
-\frac{1}{2}
\left(
\frac{\vartheta-\mu(\vartheta)}{\sigma(\vartheta)}
\right)^2
\right],
\qquad
\vartheta_{\rm min}<\vartheta<\vartheta_{\rm max},
\label{eq:gaussian_prior_qfmg}
\end{equation}
with $\pi(\vartheta)=0$ outside the allowed interval.

The final prior values obtained from this procedure are summarized in Tab.~\ref{prior}. These priors are not informed by independent mass measurements; they are obtained only from the observed QPO frequencies together with the physical requirements of horizon existence and orbital stability. The subsequent MCMC analysis therefore constrains $M$, $K$, $\alpha$, and $r$ self-consistently, separately for each of the RP, WD, and ER4 models.

\begin{table*}
\centering
\caption{Prior values.}
\label{prior}
\renewcommand{\arraystretch}{1.4}
\setlength{\tabcolsep}{5pt}
\begin{tabular}{l l c c c c c c c c}
\hline\hline
Source & Model & $\mu_M$ & $\sigma_M$ & $\mu_K$ & $\sigma_K$ & $\mu_\alpha$ & $\sigma_\alpha$ & $\mu_r$ & $\sigma_r$ \\
\hline
\multirow{3}{*}{XTE J1550$-$564} & RP & $6.7339$ & $0.1096$ & $0.2069$ & $0.1479$ & $0.4691$ & $0.0867$ & $6.7063$ & $0.0702$ \\
 & WD & $7.8946$ & $0.3113$ & $0.2155$ & $0.1386$ & $0.4511$ & $0.0947$ & $7.9002$ & $0.1834$ \\
 & ER4 & $9.0307$ & $0.0601$ & $0.2105$ & $0.1523$ & $0.4945$ & $0.0669$ & $6.2355$ & $0.0274$ \\
\hline
\multirow{3}{*}{GRO J1655$-$40} & RP & $4.0911$ & $0.0605$ & $0.2130$ & $0.1434$ & $0.4679$ & $0.0829$ & $6.7395$ & $0.0602$ \\
 & WD & $4.7470$ & $0.1510$ & $0.2153$ & $0.1435$ & $0.4478$ & $0.0983$ & $7.9673$ & $0.1606$ \\
 & ER4 & $5.5215$ & $0.0329$ & $0.2019$ & $0.1485$ & $0.4998$ & $0.0707$ & $6.2494$ & $0.0206$ \\
 \hline
\multirow{3}{*}{M82 X$-$1} & RP & $361.211$ & $5.759$ & $0.2166$ & $0.1505$ & $0.4686$ & $0.0895$ & $6.7675$ & $0.0682$ \\
 & WD & $415.358$ & $14.749$ & $0.2313$ & $0.1460$ & $0.4488$ & $0.1011$ & $8.0308$ & $0.1785$ \\
 & ER4 & $491.625$ & $3.213$ & $0.1983$ & $0.1417$ & $0.5038$ & $0.0795$ & $6.2604$ & $0.0233$ \\
\hline
\multirow{3}{*}{NGC 1313 X$-$1} & RP & $3868.54$ & $168.17$ & $0.2124$ & $0.1490$ & $0.4650$ & $0.0861$ & $6.9021$ & $0.1227$ \\
 & WD & $4278.23$ & $327.28$ & $0.2328$ & $0.1471$ & $0.4439$ & $0.0970$ & $8.3267$ & $0.2892$ \\
 & ER4 & $5437.15$ & $113.97$ & $0.2107$ & $0.1452$ & $0.4919$ & $0.0754$ & $6.3105$ & $0.0467$ \\
\hline\hline
\end{tabular}
\end{table*}

\subsection{Posterior constraints and best-fit values}

\begin{figure*}
     \centering
     \includegraphics[width=0.49\linewidth]{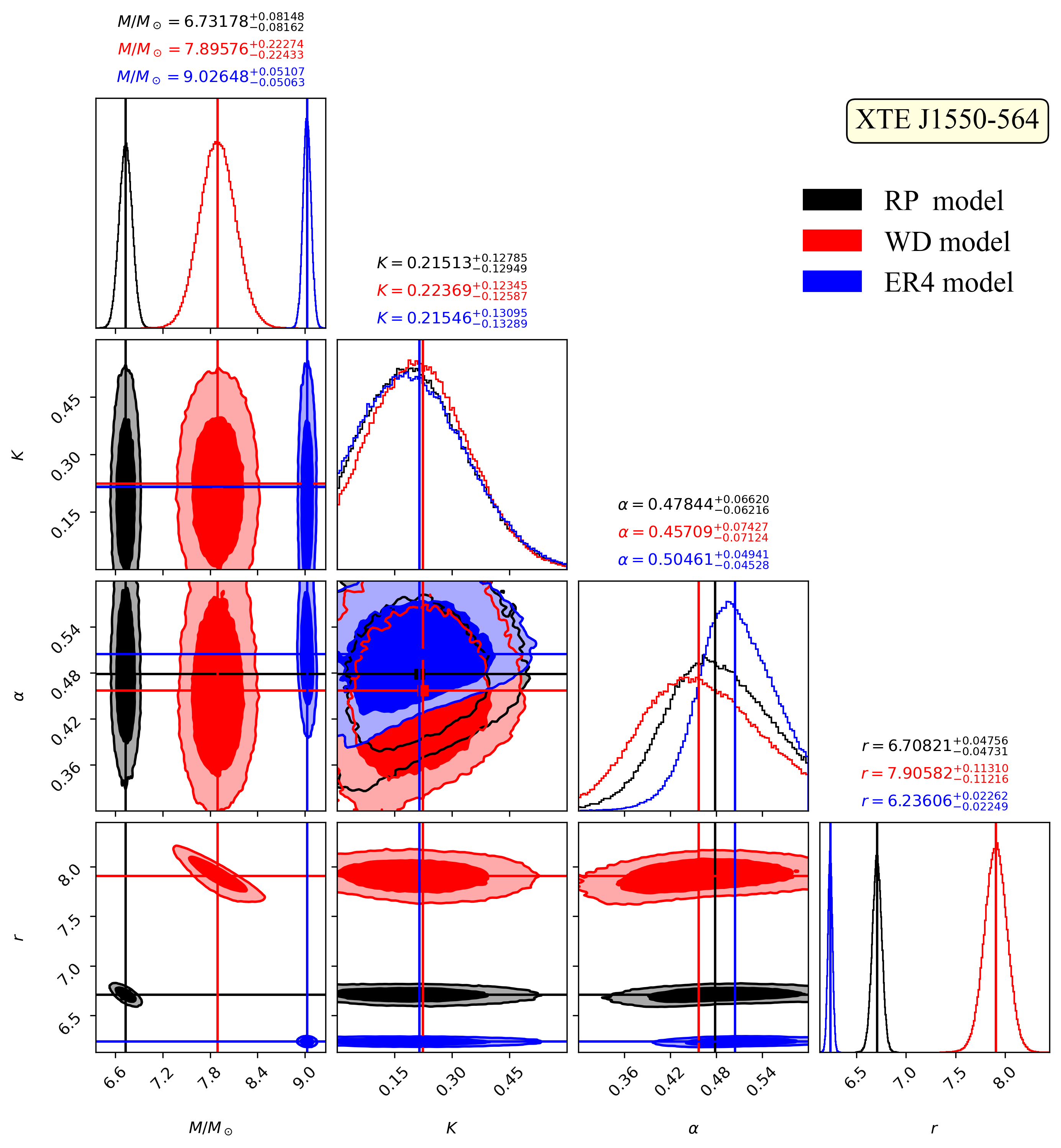}
     \includegraphics[width=0.49\linewidth]{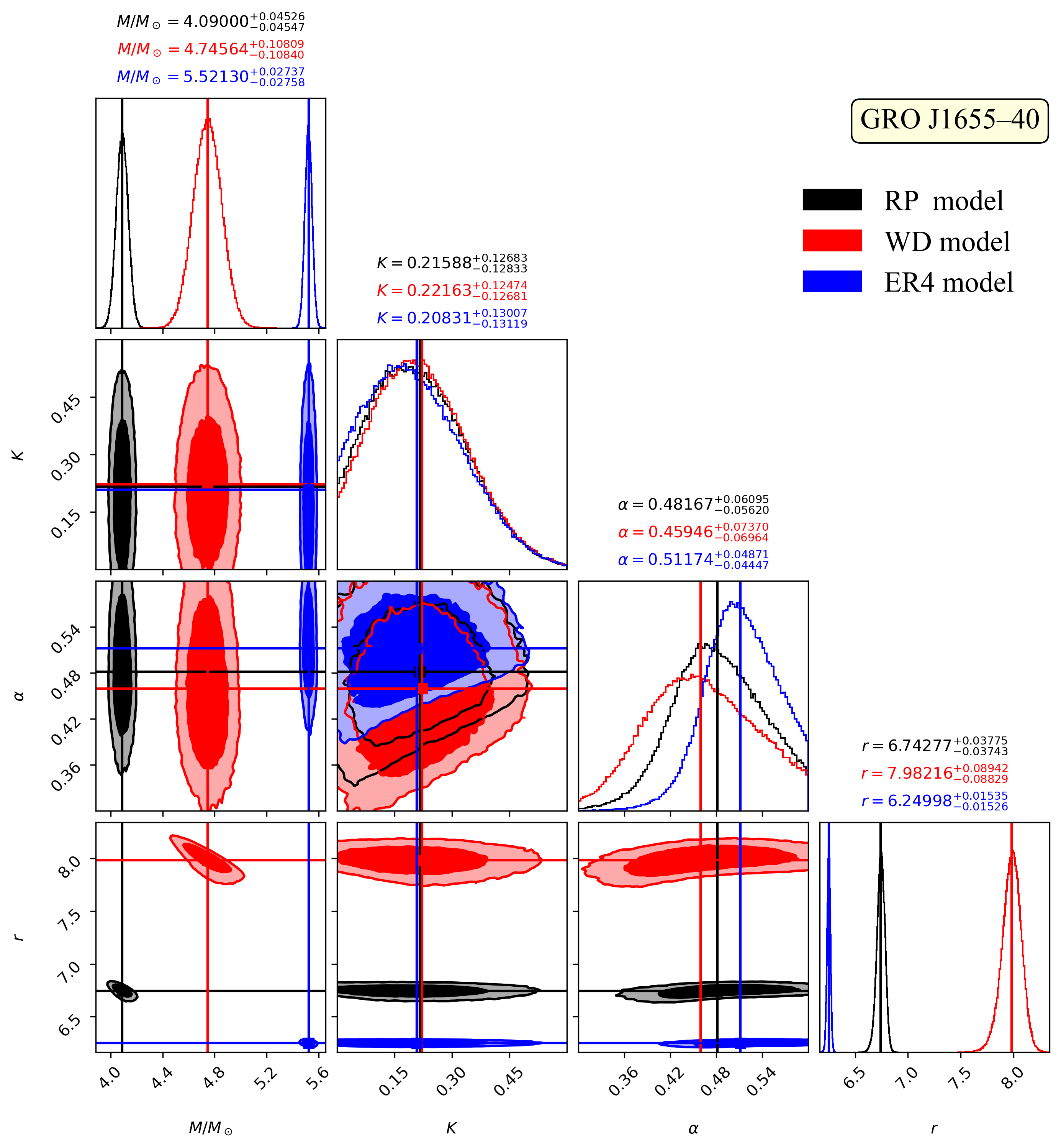}
     \includegraphics[width=0.49\linewidth]{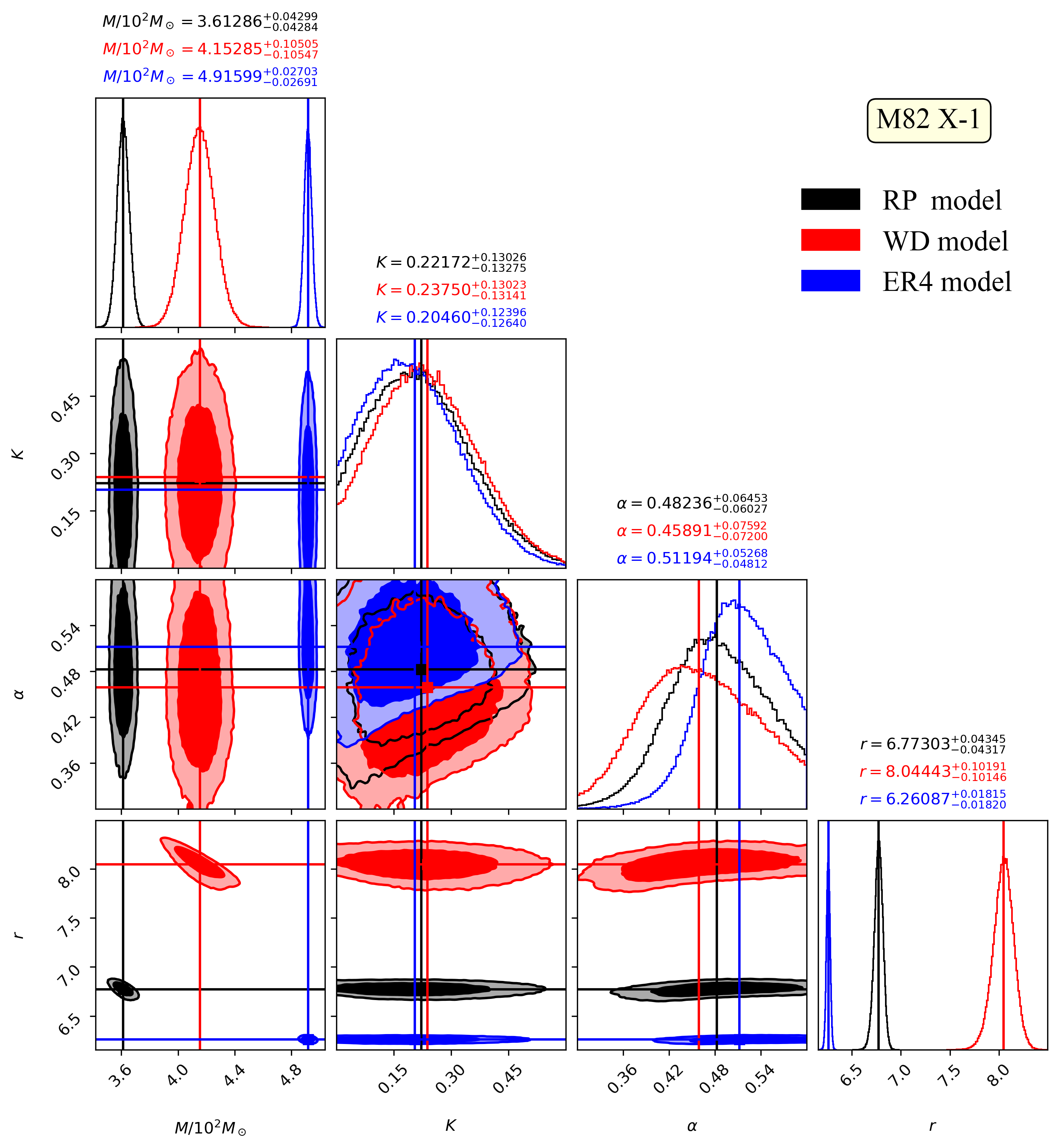}
     \includegraphics[width=0.49\linewidth]{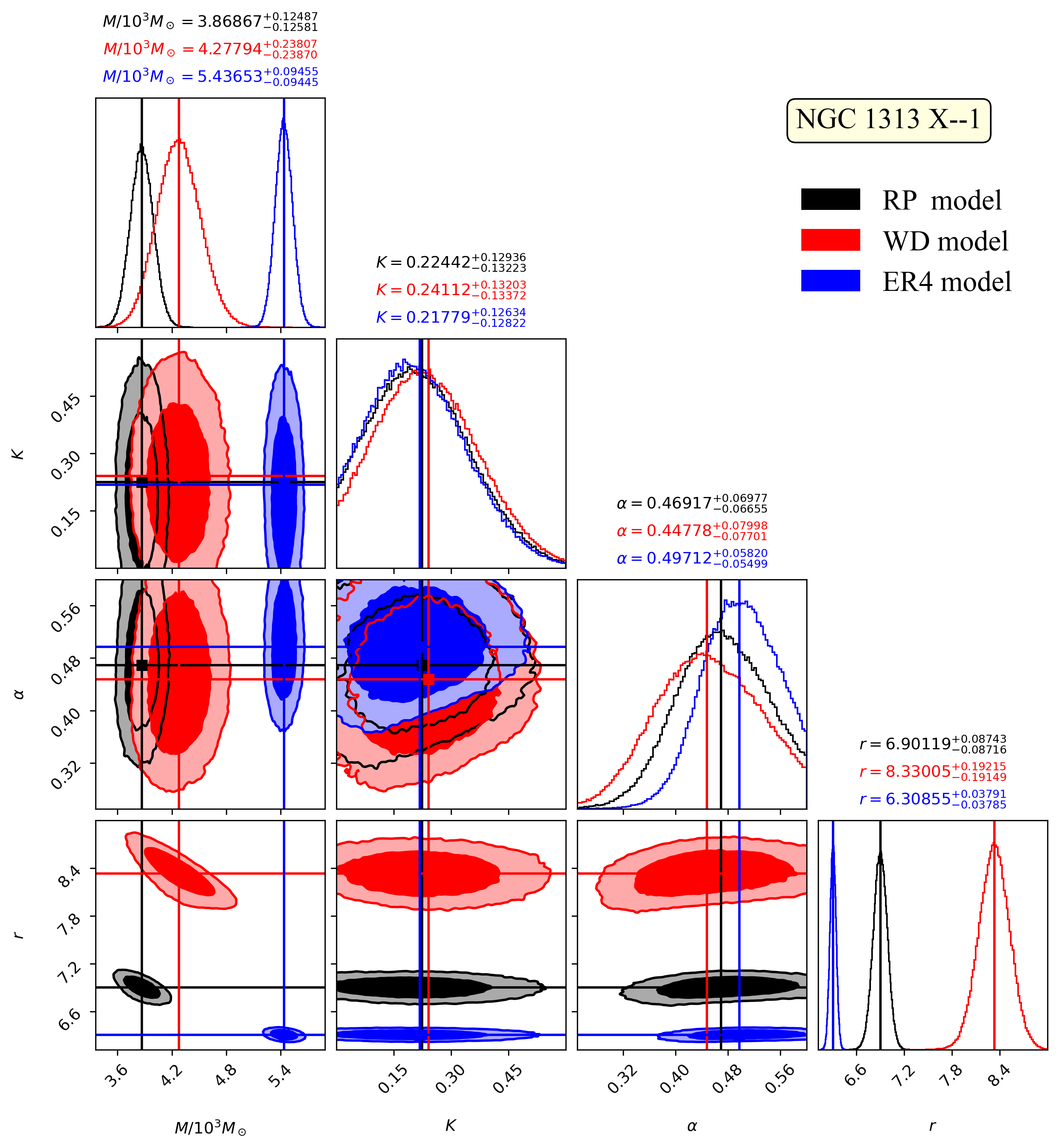}
     \caption{Posterior distributions of $M$, $K$, $\alpha$, and $r$ for the selected twin-peak QPO sources.}
     \label{fig10}
 \end{figure*}

Using the Gaussian priors of Table~\ref{prior}, we run the \texttt{emcee} sampler \citep{2013PASP..125..306F} separately for each source and each of the RP, WD, and ER4 models, checking convergence via the acceptance fraction and the integrated autocorrelation time before discarding the burn-in and thinning the remaining chains. For each parameter, we report the median of the marginalized posterior as the best-fit value, with $1\sigma$ credible intervals, while the two-dimensional posterior contours reveal correlations and degeneracies among $M$, $K$, $\alpha$, and $r$. The resulting posterior distributions and best-fit constraints are presented in Fig.~\ref{fig10} and Tab.~\ref{tab:posteriors}.

\begin{table*}
\centering
\caption{Posterior best-fit values.}
\label{tab:posteriors}
\renewcommand{\arraystretch}{1.4}
\setlength{\tabcolsep}{8pt}
\begin{tabular}{l l c c c c}
\hline\hline
Source & Model & $M/M_\odot$ & $K$ & $\alpha$ & $r$ \\
\hline
\multirow{3}{*}{XTE J1550$-$564} & RP & $6.7318^{+0.0815}_{-0.0816}$ & $0.2151^{+0.1278}_{-0.1295}$ & $0.4784^{+0.0662}_{-0.0622}$ & $6.7082^{+0.0476}_{-0.0473}$ \\
 & WD & $7.8958^{+0.2227}_{-0.2243}$ & $0.2237^{+0.1235}_{-0.1259}$ & $0.4571^{+0.0743}_{-0.0712}$ & $7.9058^{+0.1131}_{-0.1122}$ \\
 & ER4 & $9.0265^{+0.0511}_{-0.0506}$ & $0.2155^{+0.1310}_{-0.1329}$ & $0.5046^{+0.0494}_{-0.0453}$ & $6.2361^{+0.0226}_{-0.0225}$ \\
\hline
\multirow{3}{*}{GRO J1655$-$40} & RP & $4.0900^{+0.0453}_{-0.0455}$ & $0.2159^{+0.1268}_{-0.1283}$ & $0.4817^{+0.0609}_{-0.0562}$ & $6.7428^{+0.0377}_{-0.0374}$ \\
 & WD & $4.7456^{+0.1081}_{-0.1084}$ & $0.2216^{+0.1247}_{-0.1268}$ & $0.4595^{+0.0737}_{-0.0696}$ & $7.9822^{+0.0894}_{-0.0883}$ \\
 & ER4 & $5.5213^{+0.0274}_{-0.0276}$ & $0.2083^{+0.1301}_{-0.1312}$ & $0.5117^{+0.0487}_{-0.0445}$ & $6.2500^{+0.0154}_{-0.0153}$ \\
\hline
\multirow{3}{*}{M82 X$-$1} & RP & $361.286^{+4.299}_{-4.284}$ & $0.2217^{+0.1303}_{-0.1328}$ & $0.4824^{+0.0645}_{-0.0603}$ & $6.7730^{+0.0435}_{-0.0432}$ \\
 & WD & $415.285^{+10.505}_{-10.547}$ & $0.2375^{+0.1302}_{-0.1314}$ & $0.4589^{+0.0759}_{-0.0720}$ & $8.0444^{+0.1019}_{-0.1015}$ \\
 & ER4 & $491.599^{+2.703}_{-2.691}$ & $0.2046^{+0.1240}_{-0.1264}$ & $0.5119^{+0.0527}_{-0.0481}$ & $6.2609^{+0.0181}_{-0.0182}$ \\
 \hline
\multirow{3}{*}{NGC 1313 X$-$1} & RP & $3868.67^{+124.87}_{-125.81}$ & $0.2244^{+0.1294}_{-0.1322}$ & $0.4692^{+0.0698}_{-0.0665}$ & $6.9012^{+0.0874}_{-0.0872}$ \\
 & WD & $4277.94^{+238.07}_{-238.70}$ & $0.2411^{+0.1320}_{-0.1337}$ & $0.4478^{+0.0800}_{-0.0770}$ & $8.3300^{+0.1921}_{-0.1915}$ \\
 & ER4 & $5436.53^{+94.55}_{-94.45}$ & $0.2178^{+0.1263}_{-0.1282}$ & $0.4971^{+0.0582}_{-0.0550}$ & $6.3086^{+0.0379}_{-0.0379}$ \\
\hline\hline
\end{tabular}
\end{table*}

\begin{figure}
    \centering
    \includegraphics[width=0.85\linewidth]{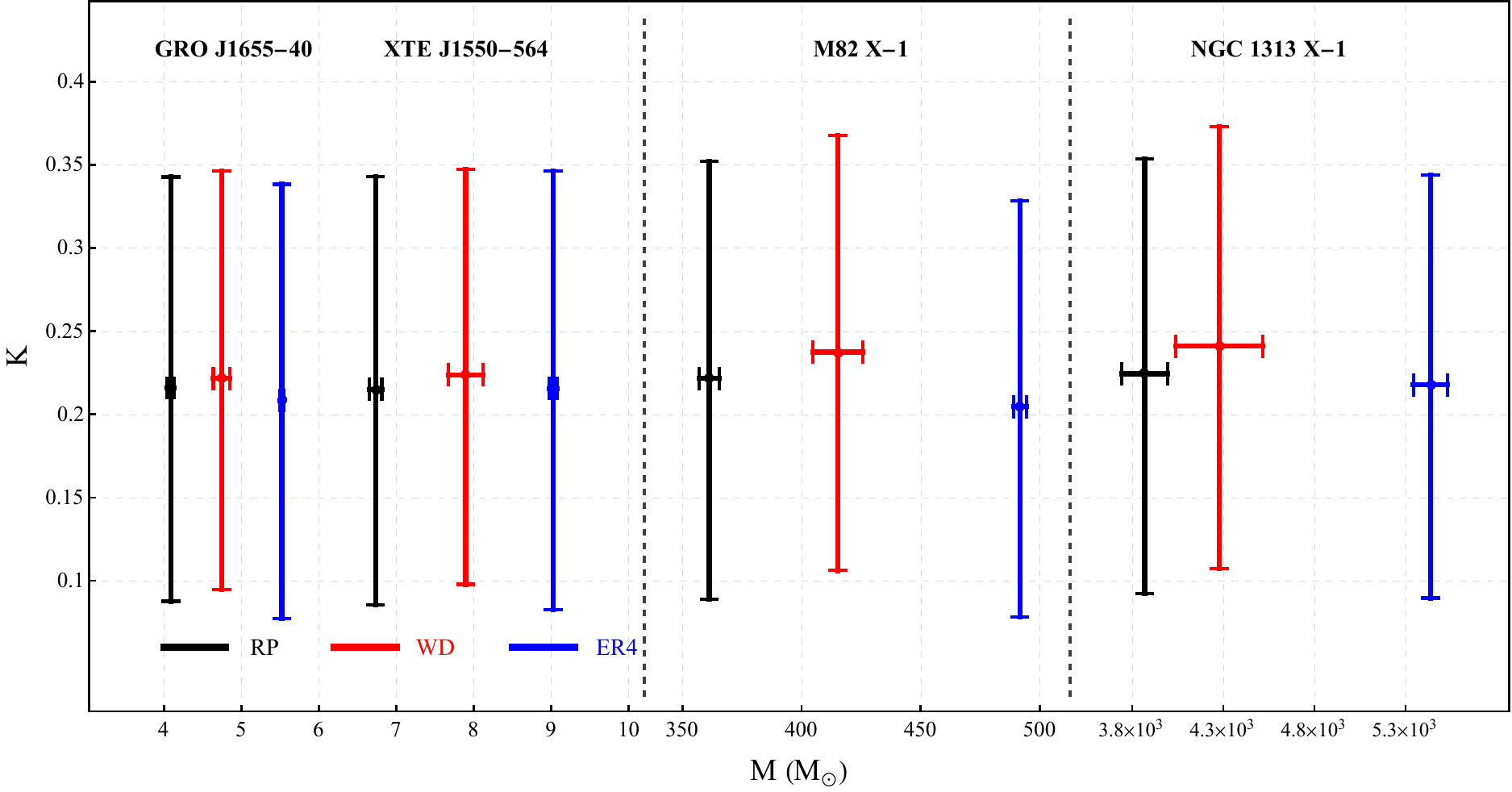}
    \includegraphics[width=0.85\linewidth]{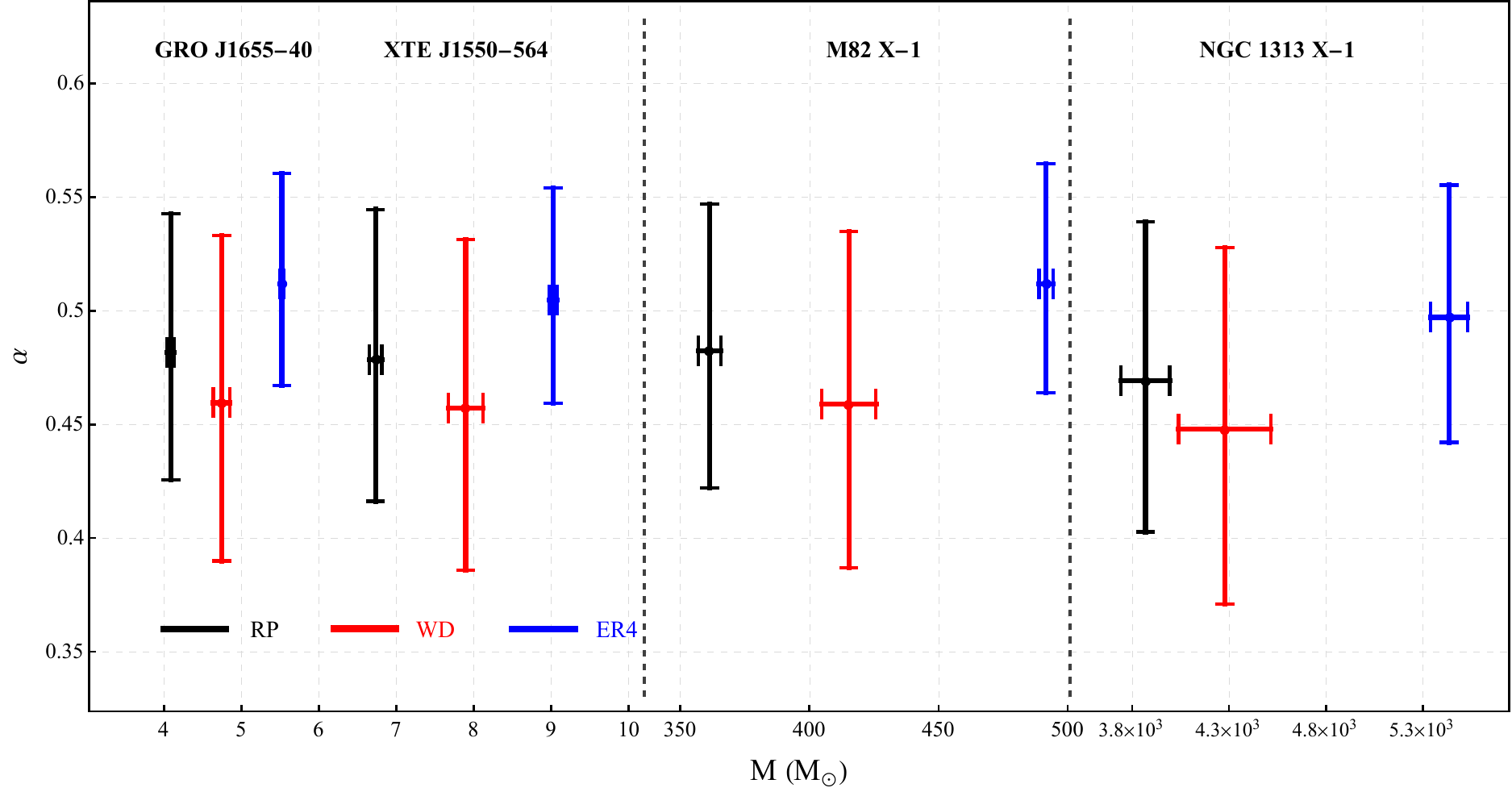}
    \caption{Posterior best-fit values of $K$ (top panel) and $\alpha$ (bottom panel) correlated with black hole mass $M$, for the RP, WD, and ER4 models, with $1\sigma$ error bars.}
    \label{f11}
\end{figure}

\begin{figure}
    \centering
    \includegraphics[width=0.245\linewidth]{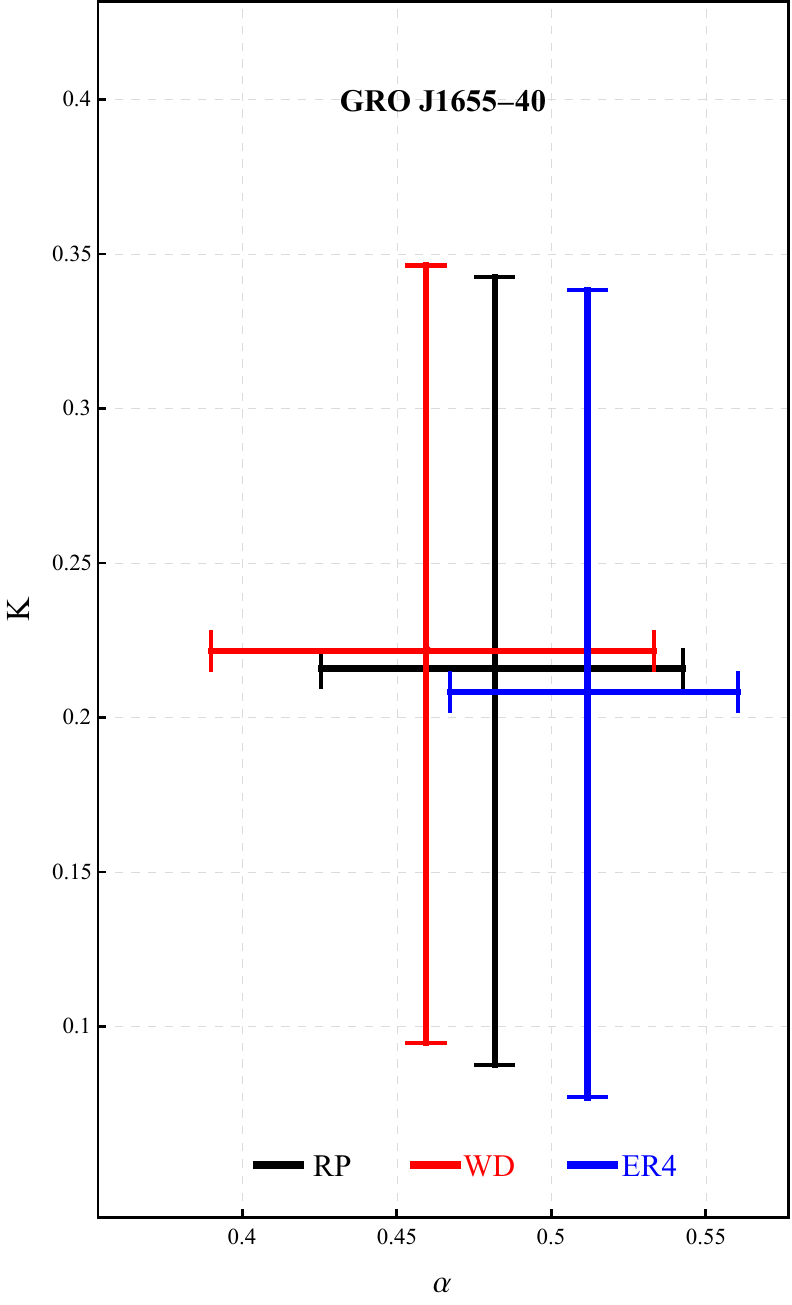}
    \includegraphics[width=0.245\linewidth]{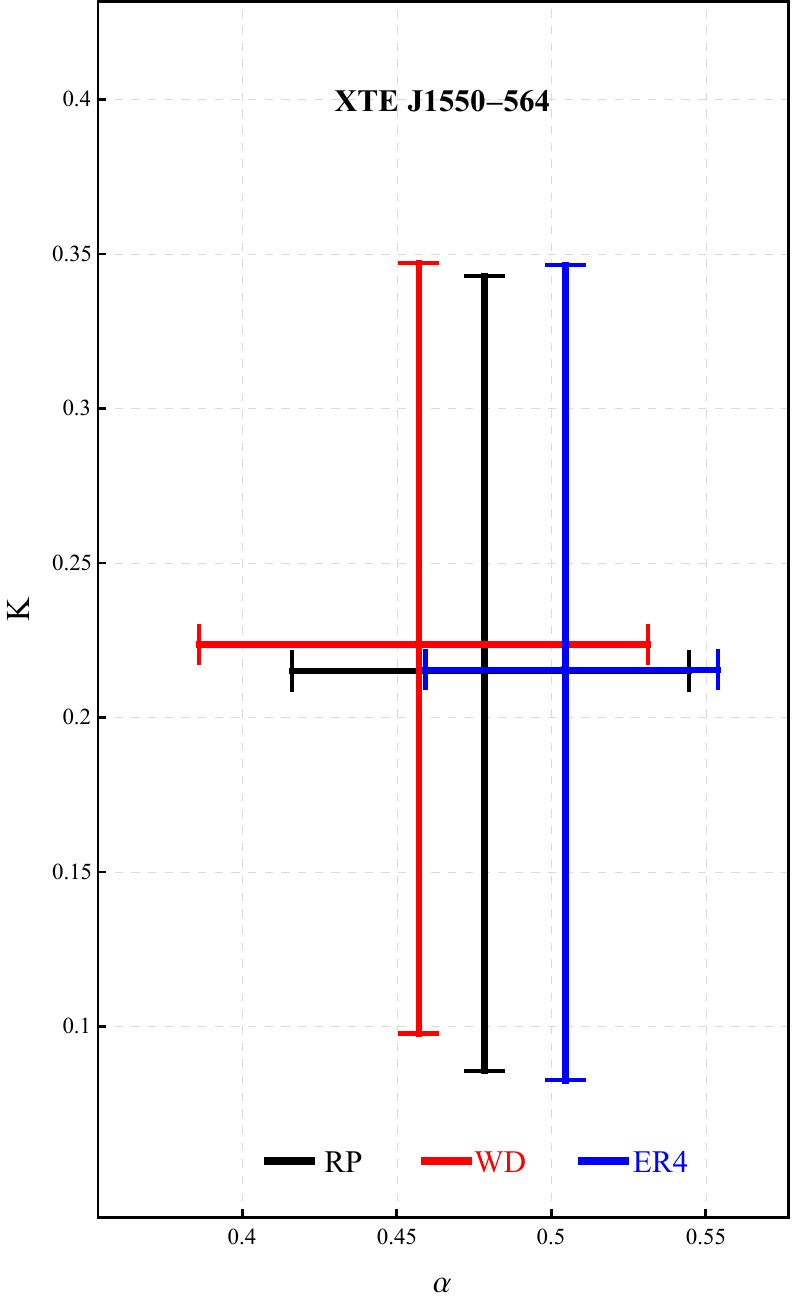}
    \includegraphics[width=0.245\linewidth]{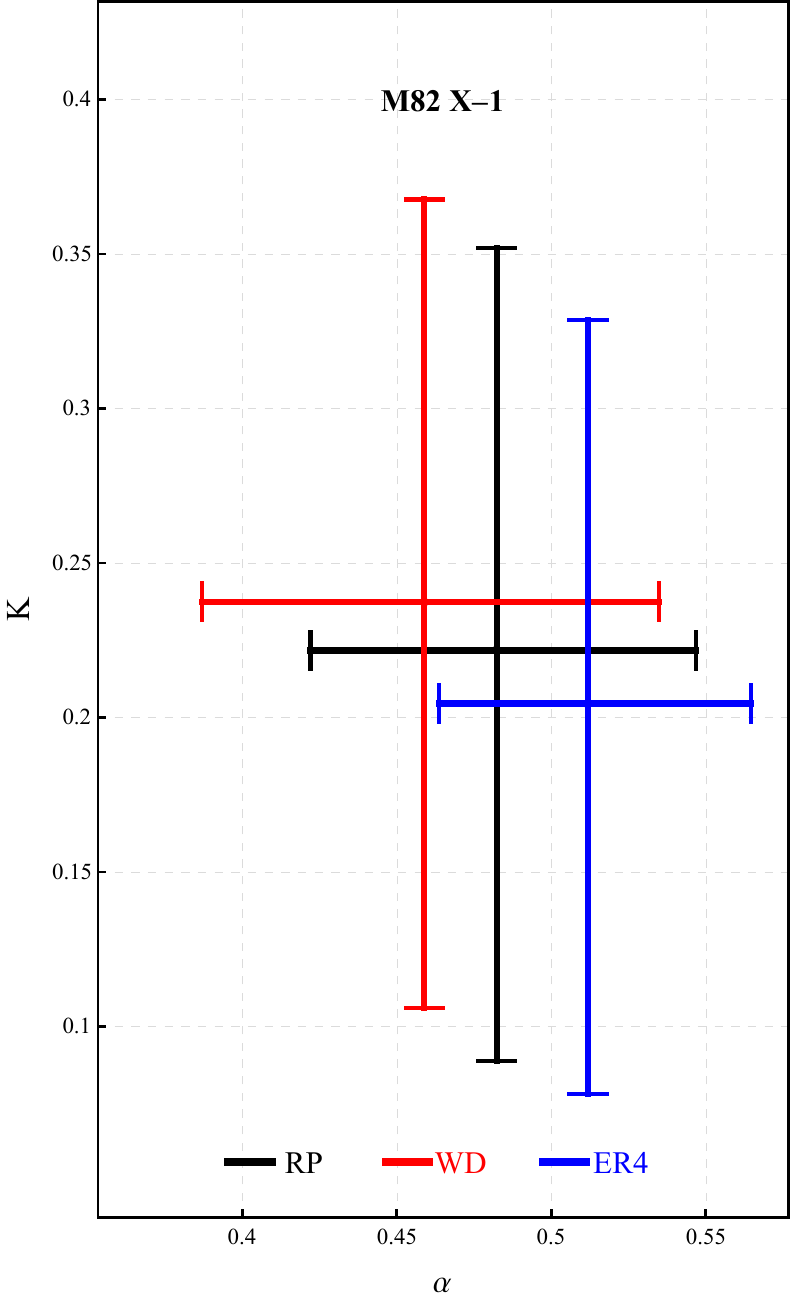}
    \includegraphics[width=0.245\linewidth]{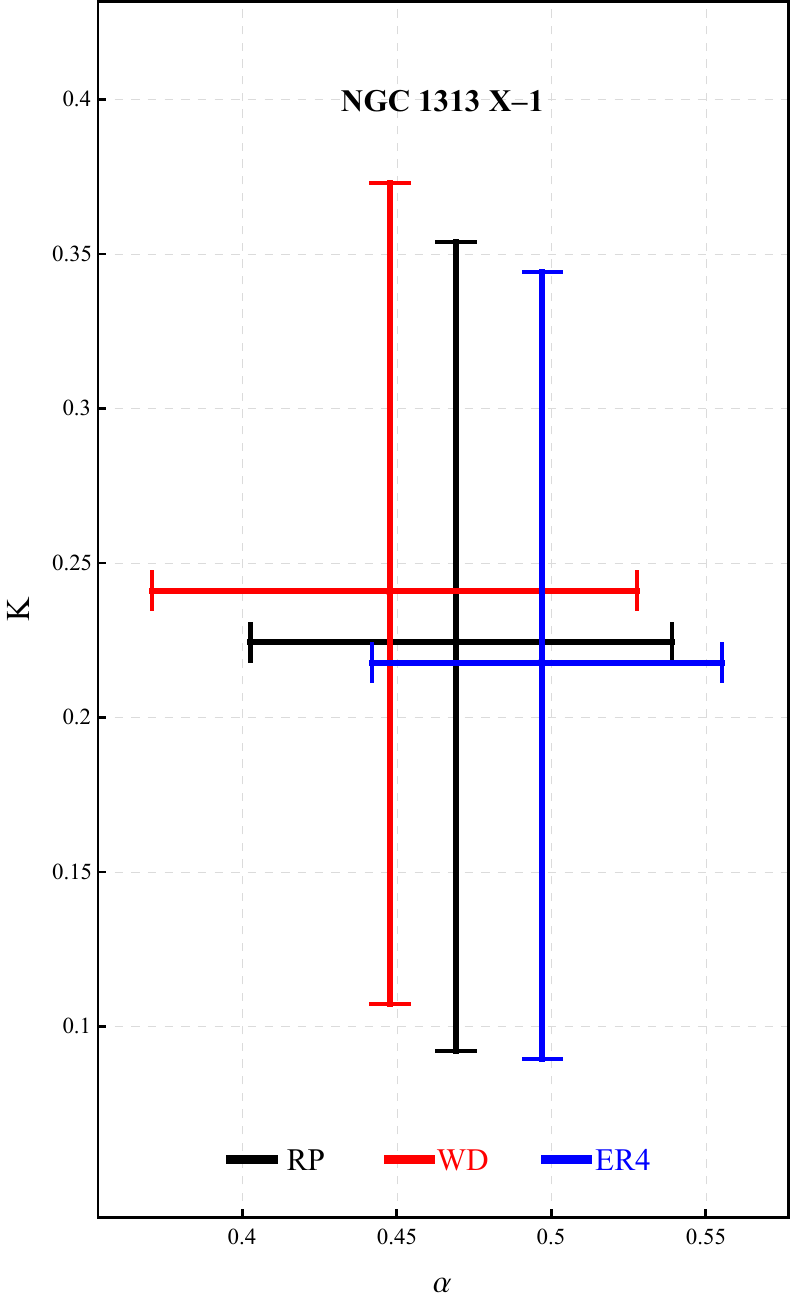}
    \caption{Posterior $K$-$\alpha$ best-fit values, with $1\sigma$ error bars, for the RP, WD, and ER4 models, shown separately for each source.}
    \label{fig:placeholder}
\end{figure}

The following conclusions can be drawn as follow:
\begin{itemize}
\item It is shown from Tab.~\ref{tab:posteriors} and Fig.~\ref{f11} that the ordering $M_{\rm RP}<M_{\rm WD}<M_{\rm ER4}$ holds strictly for all four sources, indicating that the choice of QPO model shifts the inferred black hole mass in a fixed, systematic direction rather than a random one. The size of this shift, however, grows with the black hole mass scale: for the stellar-mass sources (GRO~J1655$-$40 and XTE~J1550$-$564, $M\sim4$--$9\,M_\odot$) the overall RP-to-ER4 difference stays close to $34$--$35\%$ ($34.99\%$ and $34.09\%$, respectively), while for the intermediate-mass candidates (M82~X$-$1 and NGC~1313~X$-$1) it grows to $36.07\%$ and $40.53\%$. This growth is not uniform across the two steps of the order: splitting the difference into RP$\to$WD and WD$\to$ER4 segments shows that the RP$\to$WD step stays comparatively small and stable across sources ($10.58\%$--$17.29\%$, smallest for NGC), whereas the WD$\to$ER4 step varies far more sharply ($14.32\%$--$27.08\%$) and is largest, by nearly a factor of $1.5$--$2$, precisely for NGC; the overall growth with mass scale is therefore driven almost entirely by the WD$\to$ER4 step, with the RP$\to$WD step remaining largely independent of it.

This model reflects how tightly each model's orbital radius is pinned down by the observed frequencies. For the stellar-mass sources, the observed QPO frequencies are high ($\sim100$--$450$~Hz), which restricts all three models to a narrow radial range close to the ISCO and keeps their mass estimates close to one another; for the intermediate-mass candidates, the observed frequencies are far lower ($\sim$ Hz), so a broader range of mass-radius combinations can reproduce them, the mass-radius degeneracy strengthens, and the modest algebraic differences between the RP, WD, and ER4 frequency relations translate into a correspondingly larger spread in the inferred mass. In practice, this means that while the choice of QPO model has little bearing on mass estimates for stellar-mass microquasars, fitting the mass of an intermediate-mass black hole candidate from its QPOs requires applying more than one QPO model.

\item The relative spread among the three models, $(K_{\max}-K_{\min})/K_{\text{best}}$, shown in the top panel of Fig.~\ref{f11}, stays small for the stellar-mass sources ($6.18\%$ for GRO~J1655$-$40 and $3.94\%$ for XTE~J1550$-$564) but grows substantially for the intermediate-mass candidates ($14.87\%$ for M82~X$-$1 and $10.23\%$ for NGC~1313~X$-$1), with the same ordering $K_{\rm ER4}<K_{\rm RP}<K_{\rm WD}$ holding for all four sources -- mirroring the ordering $r_{\rm ER4}<r_{\rm RP}<r_{\rm WD}$ already found for the orbital radius in Table~\ref{tab:posteriors}. This pairing has a natural physical origin: since the dust-field term decays rapidly with radius, reproducing the same frequency deviation from Schwarzschild at the larger orbital radius favored by the WD model requires a larger $K$, whereas the smaller radius favored by ER4 reaches the same deviation with a smaller $K$. This radius-$K$ coupling stays negligible where the mass-radius degeneracy is weak, as in the high-frequency stellar-mass sources, but amplifies into a substantial spread in $K$ where the degeneracy is strong, as in the low-frequency intermediate-mass candidates. Consequently, the choice of QPO model has little bearing on the inferred value of $K$ in the stellar-mass regime, but becomes significant at the $\sim100$ and $\sim1000\,M_\odot$ scales, indicating that constraining $K$ reliably for intermediate-mass black hole candidates requires cross-checking against at least two QPO models.

\item The bottom panel of Fig.~\ref{f11} shows the correlation between $\alpha$ and $M$; the relative spread among the three models, $(\alpha_{\max}-\alpha_{\min})/\alpha_{\text{best}}$, is comparable across all four sources ($10.85\%$ for GRO~J1655$-$40, $9.90\%$ for XTE~J1550$-$564, $10.94\%$ for M82~X$-$1, and $10.46\%$ for NGC~1313~X$-$1), differing by less than one percentage point between them. This is markedly different from the behavior found for $K$ and $M$, where the spread grew by several factors between the stellar-mass and intermediate-mass sources: here the sensitivity of the inferred $\alpha$ to the choice of QPO model is essentially independent of the black hole mass scale. Moreover, in all three models and for all four sources, $\alpha$ is recovered within the narrow range $0.44$--$0.51$, coinciding with the saturated regime $\alpha\in(0.4,0.6)$ already identified in Fig.s~\ref{V.eff}, \ref{f6}, and \ref{f9}, where the sensitivity of the QF effect to $\alpha$ drops sharply even though the effect itself remains clearly distinguishable from the Schwarzschild case. The MCMC analysis therefore provides independent evidence that, irrespective of black hole mass, the observed twin-peak QPO data place the QF parameter squarely within this saturated regime, where the sensitivity of the strong-field orbital structure to $\alpha$ is at its weakest.

\item For GRO~J1655$-$40, independent mass determinations span a fairly wide range: early dynamical and photometric modeling in quiescence gave $M=7.02\pm0.22\,M_\odot$ \citep{1997ApJ...477..876O}, with inclination and binary-geometry constraints refined by \citep{1998A&A...329..538V}; subsequent optical spectroscopy narrowed the allowed interval to $5.5$--$7.9\,M_\odot$ \citep{1999MNRAS.306...89S}, while combined optical/near-infrared photometry gave $M=6.3\pm0.5\,M_\odot$ \citep{2001ApJ...554.1290G} and self-consistent ellipsoidal light-curve modeling favored a lower value, $M=5.4\pm0.3\,M_\odot$ \citep{2002MNRAS.331..351B}, also adopted in the BlackCAT catalog \citep{2016A&A...587A..61C}. QPO-based estimates cluster near this lower branch: the relativistic-precession fit to the QPO triplet gives $M=5.31\pm0.07\,M_\odot$ \citep{2014MNRAS.437.2554M}, a Kerr-background RP analysis gives $M=5.30\pm0.11\,M_\odot$ \citep{2015EPJC...75..162B}, and a systematic comparison of QPO prescriptions gives $M=5.3\pm0.1\,M_\odot$ (RP), $5.5\pm0.1\,M_\odot$ (total precession), and $5.1\pm0.1\,M_\odot$ (epicyclic resonance with beat frequency) \citep{2016A&A...586A.130S,2016ApJ...825...13S}; further modified-gravity QPO studies place the mass between this lower branch and the higher dynamical branch of $6$--$7\,M_\odot$ \citep{2021EPJC...81.1043J,2022MNRAS.517.1389R,2024EPJC...84..420A,2024PDU....4601603D,2024PDU....4601569D,2024EPJC...84.1114R,2025NuPhB101416873A}. Against this range, our ER4-model estimate, $M=5.52\,M_\odot$, falls within the lower QPO branch and close to the optical benchmark of \citep{2002MNRAS.331..351B}, while our WD estimate, $M=4.75\,M_\odot$, and RP estimate, $M=4.09\,M_\odot$, lie below essentially the full interval spanned by the independent literature.

\item For XTE~J1550$-$564, the dynamical mass function gives $M=6.86\pm0.71\,M_\odot$ and a most likely mass $M\simeq9.41\,M_\odot$ ($1\sigma$ interval $8.36$--$10.76\,M_\odot$), rising to $M\simeq10.56\,M_\odot$ ($9.68$--$11.58\,M_\odot$) once the companion's rotational velocity is included \citep{2002ApJ...568..845O}; an improved dynamical model later gave the more precise and widely adopted reference value $M=9.10\pm0.61\,M_\odot$ \citep{2011ApJ...730...75O}. Several QPO-based studies in modified gravity closely reproduce this reference value: $M=9.22^{+0.23}_{-0.25}\,M_\odot$ in Starobinsky--Bel--Robinson gravity \citep{2024EPJC...84..420A} and $M=9.13^{+0.28}_{-0.27}\,M_\odot$ in a magnetized regular black hole with a Minkowski core \citep{2025EPJC...85...95G}, while a braneworld magnetized black hole analysis gives a somewhat higher value, $M\simeq10.47\pm1.01\,M_\odot$, still partially overlapping the upper side of the dynamical interval \citep{2024EPJC...84.1114R}; other recent QPO studies adopt the dynamical value directly as their prior \citep{2024PDU....4601603D,2026JCAP...01..044X}. Our ER4-model estimate, $M=9.03\,M_\odot$, sits close to both the dynamical reference mass and the modified-gravity QPO estimates, whereas our WD estimate, $M=7.90\,M_\odot$, and RP estimate, $M=6.73\,M_\odot$, fall below the observationally allowed interval, the latter by a substantial margin.

\item For M82~X$-$1, early spectral work placed its mass in the broad intermediate range $10^2$--$10^4\,M_\odot$ based on its super-Eddington luminosity \citep{2001MNRAS.321L..29K}, while a thermal-dominant spectral-state analysis, assuming a rapidly rotating black hole, gave $M\in200$--$800\,M_\odot$ independent of any QPO information \citep{2010ApJ...712L.169F}. QPO-based estimates followed the discovery of the source's high-frequency QPOs \citep{2003ApJ...586L..61S}: spectral-timing--QPO correlations gave $M\sim10^3\,M_\odot$ \citep{2004ApJ...614L.113F}, and combining a low-frequency break with the QPO gave $M\simeq25$--$520\,M_\odot$ \citep{2006ApJ...637L..21D}. A more precise constraint followed the discovery of the stable $3{:}2$ twin-peak pair at $\nu_L=3.32\pm0.06$ and $\nu_U=5.07\pm0.06$~Hz, for which inverse mass scaling gave $M=(428\pm105)\,M_\odot$ and the RP interpretation gave $M=(415\pm63)\,M_\odot$ \citep{2014Natur.513...74P}; applying several geodesic high-frequency-QPO models to the same pair, \citet{2015MNRAS.451.2575S} found the mass strongly model dependent but bounded by $M>130\,M_\odot$, narrowing to $140$--$660\,M_\odot$ once the spin is restricted to $0.05<a<0.6$. The source has since served as a benchmark for QPO tests of magnetized \citep{2024EPJC...84.1114R}, noncommutative dark-matter \citep{2025NuPhB101416873A}, and quantum-corrected \citep{2026JCAP...01..044X} black hole models, all favoring the same intermediate-mass scale. Against this literature, our WD-model estimate, $M=415.29\,M_\odot$, coincides almost exactly with the RP-based estimate of \citep{2014Natur.513...74P}, while our RP estimate, $M=361.29\,M_\odot$, and ER4 estimate, $M=491.60\,M_\odot$, both remain within the broader bounds of \citep{2010ApJ...712L.169F} and \citep{2015MNRAS.451.2575S}.

\item For NGC~1313~X$-$1, no independent dynamical mass measurement is available, and existing constraints instead rely on spectral and QPO-scaling methods. Its cool ($kT\sim0.2$~keV) soft X-ray excess was first interpreted as thermal disk emission around a $\sim10^3\,M_\odot$ black hole \citep{2003ApJ...585L..37M}, broadly consistent with a looser spectral bound $M>100\,M_\odot$ \citep{2006AdSpR..38.1374T}; the X-ray scaling method applied across multiple epochs allowed a much wider range, $M\simeq295$--$6166\,M_\odot$ \citep{2018MNRAS.473..136J}. QPO-based evidence points more specifically to the intermediate-mass regime: the discovery of the source's $3{:}2$ twin-peak pair at $\nu_L=0.29\pm0.01$ and $\nu_U=0.46\pm0.02$~Hz, scaled against Galactic stellar-mass black holes, gave $M=(5\pm1.3)\times10^3\,M_\odot$ \citep{2015ApJ...811L..11P}, while a combined epicyclic-resonance and magnetic disk-corona fit to the same QPO pair and its X-ray spectrum gave $M=2524$--$6811\,M_\odot$ with spin $a>0.3$ \citep{2019CoSka..49....7H}. All three of our estimates, $M=3868.67\,M_\odot$ (RP), $4277.94\,M_\odot$ (WD), and $5436.53\,M_\odot$ (ER4), fall within this combined spectral-and-QPO range and overlap the $1\sigma$ interval of \citep{2015ApJ...811L..11P}, indicating good overall agreement between our twin-peak QPO constraints and the independent mass scale of this source.

\end{itemize}

\section{Conclusion}
\label{sec.7}

In this work, we have derived the black hole solution surrounded by a dust field in QFMG, characterized by the dust-field strength $K$ and the quantum-fluctuation parameter $\alpha$, and used it to study the geodesic motion, fundamental frequencies, and twin-peak QPO properties of test particles, before confronting the model with twin-peak QPO data from four observed sources -- two stellar-mass and two intermediate-mass black hole candidates -- through an MCMC analysis.

Strengthening the dust field draws the inner and outer horizons together until, for a sufficiently large $K$, they merge into a single extremal horizon, whereas strengthening the quantum-fluctuation parameter counteracts this behavior and drives the horizons back toward their Schwarzschild values.

For the effective potential and the specific energy and angular momentum of circular orbits, the dust field raises the potential barrier and shifts it closer to the black hole, while the quantum-fluctuation effect lowers the barrier and pushes it outward; both influences become negligible at large radii, consistent with the asymptotically Schwarzschild character of the spacetime.

The characteristic orbital radii -- the innermost stable circular orbit, the marginally bound orbit, and the marginal circular orbit -- are all pulled inward by a strengthening dust field and pushed outward by a strengthening quantum-fluctuation parameter, with the Schwarzschild ordering among the three radii preserved throughout the parameter space.

The Keplerian and radial epicyclic frequencies, the three QPO models' resonance radii follow the same pattern: the dust field shifts the frequency tracks above the Schwarzschild curve and draws the resonant structure inward, while the quantum-fluctuation effect acts in the opposite direction; the size of the spread among the three models, in turn, proves to be a property of the QPO model itself rather than of the spacetime.

The MCMC analysis shows that the choice of QPO model systematically shifts the inferred black hole mass, with RP giving the lowest and ER4 the highest value, and that the size of this shift grows with the black hole mass scale -- small for the stellar-mass sources and substantially larger for the intermediate-mass candidates. The same mass-dependent model sensitivity is found for $K$, but not for $\alpha$, whose spread among the three models remains comparable across all four sources, consistent with the quantum-fluctuation parameter being recovered within the same saturated regime for every black hole studied. Compared with independent mass estimates, the best-matching QPO model differs from source to source: the ER4 estimates lie closest to the independent estimates for GRO~J1655$-$40 and XTE~J1550$-$564, the WD estimate coincides almost exactly with an independent RP-based estimate for M82~X$-$1, while for NGC~1313~X$-$1 all three of our model estimates fall within the independently allowed range.

\def\prc{Phys. Rev. C}
\def\pre{Phys. Rev. E}
\def\prd{Phys. Rev. D}
\def\prl{Phys. Rev. Lett.}
\def\jcap{J. Cosmol. Astropart. Phys.}
\def\apss{Astrophys. Space Sci.}
\def\mnras{Mon. Not. R. Astron. Soc.}
\def\apj{Astrophys. J.}
\def\aap{Astron. Astrophys.}
\def\actaa{Acta Astron.}
\def\pasj{Publ. Astron. Soc. Jpn.}
\def\apjl{Astrophys. J. Lett.}
\def\pasa{Publ. Astron. Soc. Aust.}
\def\nat{Nature}
\def\physrep{Phys. Rep.}
\def\araa{Annu. Rev. Astron. Astrophys.}
\def\apjs{Astrophys. J.Suppl.}
\def\aapr{Astron. Astrophys. Rev.}
\def\procspie{Proc. SPIE}
\def\pasp{Publications of the Astronomical Society of the Pacific}

\section*{References}
\bibliographystyle{apsrev4-1}
\bibliography{main}

\end{document}